\documentstyle[epsf,epsfig]{elsart}  

\begin{document}
\newcommand{\p}{\partial}
\newcommand{\ls}{\left(}
\newcommand{\rs}{\right)}
\newcommand{\beq}{\begin{equation}}
\newcommand{\eeq}{\end{equation}}
\newcommand{\beqa}{\begin{eqnarray}}
\newcommand{\eeqa}{\end{eqnarray}}
\newcommand{\beqao}{\begin{eqnarray*}}
\newcommand{\eeqao}{\end{eqnarray*}}
\newcommand{\bdm}{\begin{displaymath}}
\newcommand{\edm}{\end{displaymath}}
\newcommand{\htau}{\hat \tau}
\newcommand{\cs}{\frac{g_{\sigma}^{2}}{m_{\sigma}^{2}}}
\newcommand{\noi}{\noindent}
\newcommand{\Gs}{\Gamma_s}
\newcommand{\Go}{\Gamma_0}
\newcommand{\Gss}{\Gamma_{\sigma}}
\newcommand{\Gv}{\Gamma_{\omega}}
\newcommand{\gs}{g_{\sigma}}
\newcommand{\gv}{g_{\omega}}
\newcommand{\ms}{m_{\sigma}}
\newcommand{\mv}{m_{\omega}}
\newcommand{\oPsi}{{\overline\Psi}}
\newcommand{\m}{{\tilde m}^*}
\newcommand{\mss}{{\tilde m}}
\newcommand{\E}{{\tilde E}^*}
\newcommand{\k}{{\tilde p}}
\newcommand{\q}{{\hat q}}
\newcommand{\alp}{{\hat a}}
\newcommand{\hj}{{\hat j}}
\newcommand{\hu}{{\hat u}}
\newcommand{\dvp}{d^{4}\!p}
\newcommand{\dvq}{d^{4}\!q}
\newcommand{\ddp}{d^{3}\!p}
\newcommand{\hpsi}{{\hat \psi}}
\newcommand{\hbpsi}{{\hat{\bar\psi}}}
\newcommand{\btau}{\mbox{\boldmath$\tau$\unboldmath}}
\newcommand{\alf}{\mbox{\boldmath$\alpha$\unboldmath}}
\newcommand{\lvecp}{\raisebox{1.77ex}{\mbox{\tiny$\leftarrow$}}
\hspace{-0.7em} \partial}
\newcommand{\rvecp}{\raisebox{1.77ex}{\mbox{\tiny$\rightarrow$}}
\hspace{-0.7em}\partial}
\begin{frontmatter}
\title{Heavy ion collisions with non--equilibrium 
Dirac--Brueckner mean fields} 
\author[muenchen]{T. Gaitanos},
\author[tuebingen]{C. Fuchs},
\author[muenchen]{H.~H. Wolter}
\address[muenchen]{Sektion Physik, Universit\"at M\"unchen, 
D-85748 Garching, Germany}
\address[tuebingen]{Institut f\"ur Theoretische Physik, 
Universit\"at T\"ubingen, D-72076 T\"ubingen, Germany}
\begin{abstract}
The influence of realistic interactions 
on the reaction dynamics in intermediate energy heavy ion 
collisions is investigated. The mean field in 
relativistic transport calculations is derived from 
microscopic Dirac-Brueckner (DB) self-energies, taking 
non-equilibrium effects, in particular the 
anisotropy of the local phase space configurations, into account. 
Thus this approach goes beyond the local 
density approximation. A detailed analysis of various 
in-plane and out-of-plane flow observables is presented for 
Au on Au reactions at incident energies ranging 
from $250$ to $800$ A.MeV and the results are compared to recent measurements 
of the FOPI collaboration. An overall 
good agreement with in--plane flow data and a reasonable description of 
the out--of--plane emission is achieved. For these results the intrinsic momentum 
dependence of the non-equilibrium mean fields is important. 
On the other hand, the local density approximation with the same underlying  
DB forces as well as a standard non--linear version of 
the $\sigma\omega$ model are less successful 
in describing the present data. This gives evidence of the applicability of 
self energies derived from the DB approach to nuclear matter also 
far from saturation and equilibrium.
\end{abstract}  
\begin{keyword} Dirac-Brueckner, non-equilibrium mean field, 
relativistic BUU, Au+Au, E=250--800 MeV/nucleon reaction, 
nucleon flow.\\
PACS numbers: {\bf 25.75.-q}, 25.60.Gc, 25.70.Mn, 
\end{keyword}
\end{frontmatter}
\section{Introduction}
Heavy ion physics at intermediate energies, i.e. up to some GeV bombarding 
energy per nucleon, open the possibility to investigate the nuclear 
equation-of-state (EOS) under extreme conditions. In contrast to the studies 
of finite nuclei which mainly yield information about the regions close 
to the saturation point of nuclear matter, in heavy ion collisions the systems 
undergoes a violent evolution from highly compressed to decompressed 
matter. However, after more than a decade 
of extensive efforts the equation-of-state is still a question of 
current debate.

On the other hand, the development of modern nuclear 
structure physics provided successful tools to handle the nuclear 
many--body problem for equilibrated systems. Especially, the relativistic 
Dirac--Brueckner Hartree--Fock (DB) approach 
\cite{hs87,thm87a,thm87b,bm90,sefu97,fuchs98} 
turned out to be a significant advance in the understanding of the 
nuclear matter saturation mechanism. This approach is 
essentially parameter free 
since it is based on a model for the bare 
NN interaction given by boson exchange 
potentials \cite{bonn89}. At present, however, because of its high complexity 
the DB approach is restricted to the description of nuclear matter 
or light nuclei. It is tempting, however, to take a 
benefit of these results also in other nuclear systems. There have 
been successful attempts to realize such a 
program with respect to the description of finite nuclei 
\cite{BoeMa94,lefu95,fule95}. Thus it is a 
natural step to continue this approach to heavy ion 
physics \cite{fgw96}.

However, ground state nuclear matter results have to be used carefully in the 
description of heavy ion reactions since here the system most of the time is 
far away from 
global or even local thermodynamical 
equilibrium \cite{essler}. As a consequence 
the local density approximation (LDA) is not well adapted to 
describe the phase space in heavy ion collisions and problems arise when 
conclusions about the equilibrium EOS are 
drawn from heavy ion collisions in this approximation. 
Instead, the anisotropy of the momentum configuration should 
be taken into account while, in contrast, the LDA refers to 
equilibrated nuclear matter, i.e. to one Fermi sphere in 
momentum space possibly with a diffuse surface if a finite 
temperature is included. Thus, the LDA only includes the 
monopole moment of the local momentum configuration and is a poor 
approximation in the case of non--equilibrium situations. One way to go 
beyond this approximation is the so called local phase space 
configuration approximation (LCA) \cite{fgw96,essler,fu92} 
which approximates the system locally 
by two Fermi spheres, i.e. by two interpenetrating currents of nuclear 
matter also called the colliding nuclear matter configuration. 
It is seen in actual calculations that 
this configuration describes fairly well the local phase space evolution 
over a large part of a heavy ion reaction \cite{essler}.

The most established models for the theoretical description of the phase 
space evolution of the 
colliding nuclei are transport theories leading to a semi-classical 
kinetic equation of a Boltzmann type  for the one-body density 
of the system, known as Boltzmann-Uehling-Uhlenbeck,
 Vlasov-\-Uehling-\-Uhlenbeck, or Landau-Vlasov 
equations (BUU, VUU, LV) \cite{gre87,bg88}. In recent years it has been found that 
even in stationary nuclear 
structure calculations a relativistic formalism 
\cite{sw86,ring96} appears to be superior to a non-relativistic 
one since the occurance of large scalar and vector fields leads to new 
mechanisms of saturation. 
Thus relativistic transport theories have been developed leading to 
the relativistic RBUU or RLV equation \cite{egg87,bkm93,mao98,fu95}. 
These approaches are formulated in the framework of a 
relativistic hadronic field theory (Quantum Hadron Dynamics, (QHD) 
\cite{sw86}) which includes baryons and mesons, and 
naturally contains large scalar and vector fields.

The derivation of transport equations usually starts from the 
Schwinger--Keldysh formalism of 
non--equilibrium Green functions and can be found, e.g., in Refs. 
\cite{mao98,dan84,btm90}. A transport equation in 
the T--matrix approximation was derived by Botermans and Malfliet \cite{btm90} 
and thus the connection to the DB theory of nuclear matter was  achieved. 
In this formulation it becomes clear that both, the mean field as well as 
the in--medium cross section are given through the T--matrix. 
To determine these quantities in a fully consistent way the equations for 
the effective interaction, i.e. the Bethe--Salpeter equation, has to be solved 
simultaneously with the kinetic equation. This, however, is a problem too complex 
to solve . However, an approximate treatment can be performed 
within the spirit of the LCA, i.e. in a stationary colliding nuclear 
matter approximation. Sehn, Fuchs and Wolter developed a procedure to 
construct approximately Dirac--Brueckner self-energies as well as in--medium cross 
sections for colliding nuclear matter, i.e for two--Fermi--ellipsoid 
momentum configurations \cite{sehn90,sehn96,cross} starting 
from a parametrisation of Dirac--Brueckner ground state results 
\cite{thm87a}. These non--equilibrium self-energies account for the 
full dependence of the specific momentum configuration and they include 
the correlations of the relativistic in-medium T-matrix.

Of course, it has been realized earlier that the reaction dynamics 
cannot be understood only in terms of a compression/decompression 
scenario (for an overview see e.g. \cite{reis97}) and that dynamic momentum 
dependent effects plays an essential role. Hence, repulsive 
momentum dependent interactions have been introduced 
phenomenologically \cite{aist86,ai91,wbc92}. Thus the description of 
intermediate energy \cite{fopi1} flow data was considerably 
improved \cite{mbc94,sahu98}. Also in--medium 
cross sections have been derived from the relativistic 
\cite{thm87b,lima93} and 
non-relativistic G--matrix \cite{alm94,jae92}. However, 
the dependence of these quantities on non--equilibrium momentum 
configurations, i.e. beyond the LDA, has been poorly investigated. The most 
complete treatment was done by the T\"ubingen 
group where fields and cross sections were calculated from a 
non-relativistic G-matrix \cite{boh89} based on the Reid-soft-core 
interaction taking also two-Fermi-sphere configurations 
into account \cite{jae92}. 
It turned out that standard observables of heavy ion collisions as, 
e.g. the transverse flow or the balance energy, react quite sensitively  
to non--equilibrium effects of such realistic forces 
\cite{khoa92b,fu95e,leh96}. A comparison with experiment 
shows a satisfying agreement in general which in some cases seems 
to be even better than with the phenomenological, e.g. Skyrme, forces 
\cite{fopi1,uma97}. However, 
the non-relativistic G-matrix calculations  
are not able to reproduce the correct saturation 
properties of nuclear matter and thus a relativistic approach 
seems preferable.

In the present work we perform a detailed comparison of RLV calculations with 
non-equilibrium Dirac-Brueckner mean fields with recent flow data 
measured by the FOPI Collaboration at GSI
\cite{reis97,fopi1,bastid97,croch1,crochet97,croch2} 
for the system $Au$ on $Au$. The 
incident energy ranges from 250 to 800 A.MeV. These data are much 
more exclusive than, e.g., the previous Steamer Chamber data \cite{flow1} 
since they are obtained with a high centrality resolution and a 
mass selection for light fragments. The energy range 
is suited in particular to test the relativistic Brueckner 
approach since the optical potential \cite{hama90} is reasonably 
reproduced up to energies around 600--800 MeV by the DB model 
\cite{thm87a,sefu97}. In Ref. \cite{fgw96} results for transverse flow for 
400 A.MeV have already been reported. These investigations are now extended 
to other energies and a wider range of observables, i.e. fragment flow, 
squeeze-out etc.

The paper is organized as follows: To set up the context of the present 
discussion and to make the work self--contained the basic steps which lead 
to the transport equation are briefly reviewed in section 2. 
In section 3 we discuss the general structure of the relativistic 
self-energy in non-equilibrium, i.e. colliding nuclear matter, 
as it can be applied in transport calculations. Section 4 presents 
more details on the models used in the present work and section 5 
gives details on the numerical implementation of this program. 
Finally in section 6 results for $Au$ on $Au$ reactions at 250, 400, 
600 and 800 A.MeV are presented for the various approaches: 
On the one hand, the DB mean field applied in the local configuration 
approximation (called DB/LCA) and 
in the local density approximation, i.e. neglecting the non--equilibrium 
features of the phase space, (called DB/LDA) to examine the influence 
of non--equilibrium effects. We also compare to 
the NL2 parametrisation of the non--linear $\sigma\omega$--model as an example 
of a widely used phenomenological force. 
Finally we draw conclusions in section 7.

\section{Transport equation}
The derivation of a kinetic equation frequently starts from the real time 
Schwinger--Keldysh formalism \cite{ms59} of Green functions. Such derivations 
have been given in refs. \cite{dan84,btm90,henn95,fau95b,knoll}. We very briefly 
rewiew the relevant steps here. One obtains a matrix 
${\underline G}$ of Green functions, (anti-) chronologically ordered Green functions 
and correlation functions $G^{>,<}$. The correlation functions 
are defined as 
$G^>(1,1^\prime)= -i <\Psi(1)\overline{\Psi} (1^\prime)>$ and 
$G^<(1,1^\prime) = i <\overline{\Psi}(1)\Psi(1^\prime)>$ using the notation 
$\eta=(t_{\eta},{\bf x}_{\eta})$. The quantity of interest 
is $G^<$ because in the limit $t_1 = t_1'$ it corresponds to a density. 
The Dyson equation in non-equilibrium is given as a matrix 
equation  
\begin{equation}
D(1,1^\prime) {\underline G}(1,1^\prime) 
= {\underline \delta}(1-1^\prime)+\int d2 
{\underline\Sigma}(1,2){\underline G}(2,1^\prime)
\quad ,
\label{Ndyson1}
\eeq
with the definition $D(1,1^\prime) \equiv (i \gamma_{\mu} \partial_{1}^{\mu} -M) 
{\underline \delta}(1-1^\prime)$. The matrix of self-energies 
${\underline\Sigma}$ in Eq. (\ref{Ndyson1}) contains all higher order 
correlations originating from the 
higher order Green functions of the Schwinger--Keldysh hierarchy. In the 
Dirac--Brueckner approach the hierarchy is truncated at the two--body level 
by summing all two--body ladder correlations in the T--matrix which obeys 
a Bethe--Salpeter equation \cite{hs87,thm87a}. The self-energy is obtained 
from the T--matrix, taking, however, the different time orderings 
into account. To determine the 
effective interaction, i.e. the T-matrix, as the solution of the 
Bethe-salpeter equation for general non-equilibrium phase space configurations 
is an extremely complex and yet unsolved problem. Thus 
suitable approximations have to be found, which is the object of the 
next section.

A kinetic equation for the correlation function $G^<$ is obtained 
by subtracting from the Dyson equation (\ref{Ndyson1}) its adjoint. A 
Wigner transformation then allows to represent the kinetic 
equation in phase space, i.e. $x-p$--space, rather than in coordinate 
space. An essential step is the truncation of the gradient 
expansion of the Wigner transform of products retaining 
only terms of first order in $\hbar$, which neglect memory terms. 
The self--energy, Eq. (\ref{Ndyson1}), is decomposed into scalar 
and vector parts 
\beq
\Sigma^{+} = \Sigma^{+}_s - \gamma^\mu \Sigma^{+}_\mu
\label{sigma1}
\eeq
and the real part of $\Sigma^{+}$ is used to define effective 
masses and kinetic momenta 
\beqa
m^* = M +  Re\Sigma^{+}_s (x,p)
\quad , \quad
p^{*}_\mu = p_\mu + Re\Sigma^{+}_\mu (x,p)
\label{pstar}
\eeqa
of the dressed particles in the nuclear medium. 
$\Sigma^{+}$ is the retarded self energy constructed by the 
difference of the corresponding correlation functions 
\linebreak
$G^{\pm}(1,1^\prime) = \theta(\pm(t_1-t_{1^\prime}))\left[G^>(1,1^\prime)
-G^<(1,1^\prime)\right]$ \cite{bm90}. The 
Dirac structure of the correlation functions $G^{>,<}$ can  
be separated off by a decomposition into a scalar spectral 
function $a$, a scalar distribution function $F$ and the projector 
onto positive energy states. In spin and isospin 
saturated systems the functions $G^{>,<}$ are then of similar form 
as in equilibrated nuclear matter \cite{hs87,thm87a} 
\beqa
G^< (x,p) &=& i \left( \not p^* + m^* \right) a(x,p) F(x,p) 
\label{NGF5}
\\
G^> (x,p) &=& -i \left( \not p^* + m^* \right) a(x,p) \left[ 1-F(x,p)\right] 
\quad .
\label{NGF6}
\eeqa

In an essential, but little investigated approximation the spectral properties 
of the baryons are 
treated in the quasiparticle approximation which is valid in the limit of a 
small imaginary part of the self energy ($Im\Sigma^+ << Re\Sigma^+$). The spectral 
function then reduces to the mass shell constraint 
$a (x,p) = 2\pi \delta \ls p^{*2} - m^{*2}\rs  2 \Theta (p_{0}^*)$ 
which sets the energy on the mass shell 
$p_{0}^* = E^* ({\bf p}) =\sqrt{ {\bf p}^{*2} + m^{*2} }$. Thus, the number of 
variables of the distribution function $F(x,p)$ is reduced from eight to seven
\beq
a(x,p) F(x,P) = 2\pi \delta [p^{*2} - m^{*2}] 2\Theta (p_{0}^* ) f(x,{\bf p}) 
\label{dist2}
\eeq
which simplifies considerably practical implementations. 

Usually the kinetic equation is treated in the Hartree approximation 
which implies to neglect the explicit momentum dependence 
of the mean field, i.e. $Re\Sigma^{+} = Re\Sigma^{+}_H (x)$. 
Then the resulting kinetic equation can be completely be 
formulated in terms of kinetic momenta 
instead of canonical momenta  \cite{bm90,egg87}
\beqa
&&  \left[p^{*\mu} \partial_{\mu}^x  + \left( p^{*}_{\nu} F^{\mu\nu}     
+ m^* \partial^{\mu}_x m^* \right) 
\partial^{p^*}_{\mu} \right] (aF) (x,p^* ) 
\nonumber\\
&=&  \frac{1}{2} \int \frac{d^4 p_{2}}{(2\pi)^4} \frac{d^4 p_{3}}{(2\pi)^4}
             \frac{d^4 p_{4}}{(2\pi)^4} 
             a(x,p)  a(x,p_2)  a(x,p_3)  a(x,p_4) W(pp_2|p_3 p_4) 
\nonumber \\    
&\times& (2\pi)^4 \delta^4 \left(p + p_{2} -p_{3} - p_{4} \right)    
\left[ F(x,p_3) F(x,p_4) \ls 1-  F(x,p) \rs \ls 1- F(x,p_2) \rs \right.  
\nonumber \\  
&-&   \left. F(x,p) F(x,p_2) 
\ls 1-  F(x,p_3) \rs \ls 1- F(x,p_4) \rs \right]    
\quad .
\label{TP5}
\eeqa
Eq. (\ref{TP5}) resembles the well known transport equation of a 
Boltzmann--Uehling--Uhlenbeck type. The left hand side is a drift term driven  
by the mean field via the kinetic momenta $p^{*}$, the field strength tensor 
$F^{\mu\nu} (x) = \p^{\nu}_x  Re\Sigma^{+\mu}_H (x) 
                -\p^{\mu}_x  Re\Sigma^{+\nu}_H (x) $, 
and the effective mass $m^*$. The right hand side is a collision integral which 
contains the transition rate $W$ or equally the in--medium cross section given by 
$(p^* + p_{2}^*)^2 \frac{d\sigma}{d\Omega}(p,p_2) = W(pp_2|p_3 p_4)$. 
As discussed in the introduction in a fully consistent approach the later is 
given by the square of the non--equilibrium T--martix \cite{bm90,cross,jae92}. 
However, in the present work we concentrate on non-equilibrium features of 
the mean field and apply a phenomenological parametrisation of the 
cross section \cite{cug81}.

\section{Colliding nuclear matter approximation}
Here we discuss approximations to construct the non-equilibrium 
self-energy for the two Fermi--sphere or colliding nuclear matter system \cite{sehn96}. 
This will be used in the transport calculations in the Local (phase space) 
Configuration Approximation (LCA). 
In this section we only consider the structure of the self-energy 
in colliding nuclear matter which is independent 
of the particular choice of the nuclear interaction.
\subsection{Local density approximation}
The mean field is commonly determined in the local density 
approximation, i.e. it is taken as that of ground state nuclear matter 
(n.m.) at the respective total density
\beqa
Re \Sigma^+ (x,p) &=& Re \Sigma^{({\rm n.m.})} \ls p_F (x), p \rs
\label{sigma_LDA}
\\
 p_F (x) &=& \ls \frac{3}{2} \pi^2 \rho_{\rm B} (x) \rs^\frac{1}{3}
\\
 \rho_{\rm B} (x) &=& 4\int \frac{d^3 p}{(2\pi)^3} f(x,{\bf p}) 
\label{rho_LDA}
\quad .
\eeqa
Here the mean field $\Sigma^{({\rm n.m.})}$ has been taken 
from different sources. 
In non-relativistic applications Skyrme forces are used 
\cite{aist86,khoa92b} or the mean field has been obtained 
microscopically from the 
G--matrix \cite{boh89,khoa92b}. In relativistic treatments, 
Eq. (\ref{TP5}), the mean field 
is usually determined in a Hartree approximation in the framework of the 
$\sigma\omega$--model \cite{sw86}, in particular of its 
non--linear extensions \cite{bkm93,BB,fu96b,koli96} 
Calculations with momentum dependent fields have been performed in Refs. \cite{mbc94} 
with self-energies of Hartree-Fock form \cite{wbc92} 
and fitted to the energy 
dependence of the empirical nucleon-nucleus optical potential \cite{hama90}. 
A similar procedure is applied in non-relativistic calculations 
when momentum dependent Skyrme forces are used \cite{aist86}. 
However, all these represent the mean field of equilibrated nuclear matter.
\subsection{Local configuration approximation}
In the {\em local (phase space) configuration approximation} (LCA) the self 
energies are parameterized for a phase space configuration of two 
inter-penetrating currents of nuclear matter, so called colliding 
nuclear matter. The covariant momentum 
distributions of the currents are given by Fermi ellipsoids 
\cite{sehn96,cross} and shown in Fig. 1. A unique 
parametrisation of such a configuration is based on 
five Lorentz invariants which are naturally chosen as 
the Fermi momenta of the currents and their relative velocity. 
In a transport calculation these parameters 
are determined from the vector currents which 
are obtained from a decomposition of 
the phase space distribution into contributions from 
projectile (1) and target 
(2), i.e. $ f^{(12)} =f^{(1)} + f^{(2)}$. 
The Fermi momenta are defined in the rest frames of 
the respective currents by 
the invariant rest densities $\rho_{0}^{(i)}$ 
\beqa
j_{\mu}^{(i)} (x) &=& 4\int \frac{d^3 p}{(2\pi)^3} 
\frac{ p_{\mu}^*}{E^*} f^{(i)} (x,{\bf p})  
\quad ,\quad i=1,2
\label{current_LCA}
\\
\rho_{0}^{(i)} (x) &=& \sqrt{ j_{\mu}^{(i)}  (x) j^{ (i)\mu}(x) }
\quad .
\label{rho_LCA}
\eeqa
The current four--velocities $u_{\mu}^{(i)} = (u_{0}^{(i)},{\bf u}^{(i)})$ 
and the relative velocity $ {\bf v}_{{\rm rel}}$ are defined as 
\beqa
u_{\mu}^{(i)} (x) &=&  j_{\mu}^{(i)} (x)/\rho_{0}^{(i)} (x)
\label{u_LCA}
\\
{\bf v}_{{\rm rel}} (x) &=& \frac{u_{0}^{(2)} {\bf u}^{(1)} - u_{0}^{(1)} 
{\bf u}^{(2)} }  {u_{\mu}^{(1)} u^{(2)\mu} }
\label{vrel_LCA}
\quad .
\eeqa
One should note that the relative velocity, Eq. (\ref{vrel_LCA}), is not 
a space-like vector but each component remains invariant under Lorentz 
transformations. In any reference frame, e.g. the 
center-of-mass (c.m.) frame of the currents, one component of 
$ {\bf v}_{{\rm rel}}$ is sufficient to characterize the configuration 
and the remaining two parameters are used to fix the reference frame. Thus 
in the following we write simply $v_{{\rm rel}}$ and refer to the 
c.m. frame of the currents. The local density approximation is recovered 
in the limit of a vanishing relative velocity 
$LCA \stackrel{v_{{\rm rel}} \to 0}{\longrightarrow} LDA$. 
 
In the LDA, instead, one would obtain the density and the 
streaming velocity of the total system (index 12) as 
\beqa
\rho_{0}^{(12)} = \sqrt{ j_{\mu}^{(12)} j^{\mu (12)}} 
=\rho_{B}^{(12)}|_{{\rm c.m.}}
\quad , \quad 
\label{rho_0_12}
 j_{\mu}^{(12)} =  u_{\mu}^{(12)} \rho_{0}^{(12)}
\quad .
\eeqa
In the c.m. frame which is the natural reference frame in colliding 
nuclear matter the spatial components vanish for the total vector current 
$j_{\mu}^{(12)} = (\rho_{0}^{(12)},{\bf 0})$ and for the
total streaming velocity $u_{\mu}^{(12)} = (1,{\bf 0})$. 
$\rho_{0}^{(12)}$ represents the the total c.m. baryon density, 
which is not the sum of the rest densities 
$\rho_{0}^{(1)}+ \rho_{0}^{(2)}$ of the currents but is larger because of 
Lorentz contraction. Therefore, in a naive LDA, Eqs. (\ref{sigma_LDA}--\ref{rho_LDA}),
the Lorentz contraction would be interpreted as a compression and could give 
misleading conclusions on the EOS. Thus a better approach in the LDA 
should be based on a total density given in terms of the rest densities.

The LCA should be able to adequately represent the time evolution of the phase 
space in a heavy ion reaction starting from the initial  
configuration of two separated cold Fermi ellipsoids, and ending 
up, possibly, with a thermalized fireball which corresponds to a single hot 
Fermi sphere. To give an impression on the quality of this representation 
we compare in Fig. \ref{phasespace_graph} the LCA parametrisation 
to the phase space 
distribution obtained from the transport calculation, 
for a central (b=0) $Au$ on $Au$ reaction at 600 A.MeV for the NL2 parameter 
set. Here we show the local momentum space in the 
central region, i.e. around ${\bf x} = 0$, at three different time steps 
($t$=5, 25 and 50 fm/c) which represent important stages of the reaction. 
The first configuration ($t$=5 fm/c) corresponds to the initial 
phase of the reaction where the nuclei start to overlap. 
The local momentum space still resembles the 
asymptotic initial configuration, 
i.e. it is given by two relatively sharp and well separated Fermi 
ellipsoids. In the compression phase ($t$=25 fm/c) where the density is 
maximal a two--Fermi--ellipsoid configuration is 
still visible. However, the ellipsoids now strongly overlap and 
due to binary collisions their shapes become more 
and more diffuse. The expansion of the system 
leads to increasing equilibration and 
the momentum configuration finally evolves 
to one single Fermi 
sphere ($t$=50 fm/c). At this stage the LDA would be appropriate. 
In the right column of Fig. \ref{phasespace_graph} a representation of the  
momentum distributions in the LCA are shown. The parameters of 
the configuration, i.e. the rest densities and the relative velocity of the 
currents are extracted from the phase space distribution of the transport calculation 
(left column) according to Eqs. (\ref{current_LCA}--\ref{vrel_LCA}). 
We then show in the right column of Fig. \ref{phasespace_graph} 
a parametrization in terms of two Fermi--distributions of finite temperature with the 
given densities and relative velocity, taking also into account the Pauli--principle 
in the overlapping part. Details are given in ref. \cite{essler}. 
This shows that a more realistic description of the momentum configuration  
requires the inclusion of finite temperatures into the 
formalism, i.e. to replace sharp Fermi ellipsoids by diffuse covariant 
Fermi distributions. This is not done in the present implementation, but 
here sharp Fermi ellipsoids are used. 

\subsection{Self--energy}
In the LCA the real part of the non--equilibrium self-energy $\Sigma^+$ 
is approximated by the self-energy in colliding nuclear matter with the 
corresponding parameters 
\beq
Re \Sigma^+ (x,p) = 
Re \Sigma^{(12)} \ls p_{F_1} (x), p_{F_2} (x), v_{{\rm rel}} (x),p\rs
\qquad ,
\label{sigma_LCA_1}
\eeq
taken in the Hartree approximation. 
To evaluate the self-energies, Eqs. (\ref{sigma_LCA_1}), 
they are expressed in terms of Lorentz invariants as for  
ground state nuclear matter \cite{hs87,thm87a,sefu97}. 
These are naturally determined in the c.m. system. 
As in nuclear matter the self-energy can be decomposed into scalar and 
vector self energies and the latter into time-like and space-like 
components \cite{sehn96,cross}
\beq
\Sigma^{(12)\mu} (p) = \Sigma_{0}^{(12)} u^{(12)\mu}  
+ \Sigma^{(12)}_{v} (p) \Delta^{(12) \mu\nu} p_\nu
\label{sigma12_2}
\eeq
with $ \Delta^{(12)\mu\nu} = g^{\mu\nu} - u^{(12)\mu} u^{(12)\mu}$ 
the projector perpendicular on the total streaming velocity 
$u^{(12)}$. The invariant functions 
$\Sigma^{(12)}_{s,0,v}$ are obtained by covariant projections 
\beqa
\Sigma^{(12)}_s &=& \frac{1}{4} tr[ \Sigma^{(12)}] 
\label{sigma_s_proj12}
\\
\Sigma^{(12)}_{0} &=& \frac{-1}{4} tr[ u^{(12)}_\mu \gamma^\mu \Sigma^{(12)}] 
\label{sigma_0_proj12}
\\
\Sigma^{(12)}_v &=& \frac{-1}{4\ls  \Delta^{(12)\mu\nu} p_\mu p_\nu \rs } 
tr\left[ \Delta^{(12)\mu\nu} p_\mu \gamma_\nu \Sigma^{(12)} \right] 
\quad .
\label{sigma_v_proj12}
\eeqa

To obtain the self-energies in a Hartree form, i.e. momentum independent, 
which is more convencient for the application in transport 
calculations, Eq. (\ref{TP5}), these are 
averaged over the total momentum configuration \cite{sehn96}. In practice we  
restrict to symmetric systems, i.e. $p_{F_1} = p_{F_2}$, such that 
the $\Sigma^{(12)}_v$ vector part vanishes which simplifies 
the task considerably. The Hartree self-energy 
takes the form 
\beq
Re\Sigma^{(12)}_H   = Re{\overline \Sigma}^{(12)}_s 
                   - \gamma_\mu Re{\overline \Sigma}^{(12)}_{0} u^{(12)\mu}  
\quad ,
\label{sigma12_2H}
\eeq
where the averaged Lorentz invariants 
(\ref{sigma_s_proj12},\ref{sigma_0_proj12}) are 
obtained as 
\beqa
 Re{\overline \Sigma}^{(12)}_s &=& 
\int \frac{\dvp}{(2\pi)^4} 
Re\Sigma^{(12)}_{s} (p)  tr\left[G^{<(12)} (p)\right] 
\bigg/ 
\int \frac{\dvp}{(2\pi)^4} tr\left[G^{<(12)} (p)\right]
\label{sigma12_s_3}
\\
 Re{\overline \Sigma}^{(12)}_{0} &=& 
\int \frac{\dvp}{(2\pi)^4} Re\Sigma^{(12)}_{0} (p) 
        tr\left[ u^{(12)}_\mu \gamma^\mu G^{<(12)} (p)\right] 
\bigg/
\int \frac{\dvp}{(2\pi)^4} tr\left[ u^{(12)}_\mu 
\gamma^\mu G^{<(12)} (p)\right]
\quad .
\label{sigma12_3}
\eeqa
As in \cite{sehn96} vertex functions are derived from the invariants in 
the following way 
\beqa
 {\overline \Gamma}^{(12)}_s =
- \frac{Re{\overline \Sigma}^{(12)}_{s} }{\rho^{(12)}_{s}} 
\quad , \quad 
 {\overline \Gamma}^{(12)}_{0} = 
 - \frac{Re {\overline\Sigma}^{(12)}_{0} }{\rho^{(12)}_0} 
\quad .
\label{gamma12_2}
\eeqa
The Hartree self-energy in symmetric colliding nuclear 
matter is now given as  
\beq
Re \Sigma_{H}^{(12)} \ls p_{F_i}, v_{{\rm rel}} \rs 
                     = -{\overline \Gamma}^{(12)}_s  
\ls p_{F_i}, v_{{\rm rel}} \rs \rho_{s}^{(12)} 
          + \gamma_\mu  {\overline \Gamma}^{(12)}_{0}  
\ls p_{F_i}, v_{{\rm rel}} \rs j^{(12)\mu}
\quad .
\label{sigma_12_H}
\eeq
It is of the same structure as in the conventional 
$\sigma\omega$--model \cite{sw86}, however the coupling constants 
for the scalar $\sigma$-- and the vector $\omega$--meson are replaced 
by vertex functions
\beqa
\frac{g_\sigma^2}{m_\sigma^2} \longmapsto 
{\overline \Gamma}^{(12)}_s(p_{F},v_{rel})   
\quad ,\quad 
\frac{g_\omega^2}{m_\omega^2} \longmapsto 
{\overline \Gamma}^{(12)}_0(p_{F},v_{rel})
\quad ,
\eeqa
which now contain the information on the anisotropy 
of the actual momentum space configuration. 

In the Dirac--Brueckner approach the invariants, Eqs. 
(\ref{sigma_s_proj12}--\ref{sigma_v_proj12}), are given in terms of the T--matrix 
\cite{hs87,thm87a,sefu97} and, in particular, 
in non--equilibrium by 
the respective non--equilibrium T--matrix. 
However, the T--matrix in colliding 
nuclear matter is not available at present. 
In \cite{sehn96} a procedure has been 
proposed to construct $\Sigma^{(12)}_{s,0,v}$ from a parametrisation of the ground state 
self energies and an extrapolation to the colliding nuclear matter system \cite{sehn96}. 

\section{The effective fields}
In the present work we compare mean fields derived from the microscopic 
Dirac-Brueckner approach to a phenomenological interaction, 
namely the non-linear $\sigma\omega$-model. In Fig. \ref{EOS_graph} 
the equations-of-states, i.e. for ground state nuclear matter, are shown 
for the DB model \cite{thm87a} and a hard (NL3) and a soft (NL2) version 
of the non--linear $\sigma\omega$--model \cite{bkm93}. 
The Dirac-Brueckner EOS is relatively soft and 
thus comparable to NL2 which is the reason why we mainly compare to 
this model. In Tab. \ref{sat_tab} the corresponding 
nuclear matter bulk properties are shown: The saturation density, 
the binding energy, the compression modulus and 
the value of the effective mass $m^*$ at saturation density. 
As discussed, e.g., in \cite{fu96b} the effective mass is a useful 
quantity to characterize the repulsiveness of the relativistic mean field 
which becomes less repulsive with increasing $m^{\ast}$. 
The DB model used here yields a small effective mass ($m^{\ast}/M=0.586$). 
However, the explicit momentum dependence of the self energy also influences 
the behavior with density, such that the DB EOS is similar to the one of 
NL2 ($m^{\ast}/M=0.83$). 

In the following the DB model is applied in two approximations, 
in the Local Density Approximation (LDA) and in the Local Configuration 
Approximation (LCA) 
where the latter accounts for the non-equilibrium features of the 
process. The self-energy which enters into the transport equation 
(\ref{TP5}) in both cases is used in the Hartree form given by Eq. 
(\ref{sigma_12_H}). In the LDA the dynamical vertex functions 
$\Gamma_{s,0} (p_{F})$ which parameterize the mean field 
depend only on the total density. 
Then the treatment is the same as in the case of finite nuclei 
\cite{lefu95,fule95} (however, rearrangement terms are not taken into 
account here.) 

In the LCA the vertex functions depend on the 
subsystem densities of the nuclear matter currents and their 
relative velocity $v_{{\rm rel}}$ (\ref{vrel_LCA}). In the present 
calculations we have restricted the determination of the 
colliding nuclear matter mean fields 
to symmetric systems, i.e. $p_{F_1} = p_{F_2}$ which we take as 
the mean value of the subsystem densities. The dynamical vertex 
functions ${\overline \Gamma}^{(12)}_{s,0}$ for colliding nuclear 
matter are determined as described in Ref. \cite{sehn96}. Thus, they include 
exchange and correlations effects of the in-medium T--matrix 
and therefore the most relevant dynamical effects of the non-equilibrium 
phase space. Of course, the approach of \cite{sehn96} 
only approximates  the solution of the 
full problem, i.e. the solution of the Bethe-Salpeter equation for 
arbitrary anisotropic configurations. The 
approach neglects the non-equilibrium effects originating 
from the configuration dependence of the Pauli operator in the 
intermediate states of the Bethe--Salpeter equation, 
which should, in principle, 
be evaluated for two-Fermi-ellipsoid configurations \cite{sehn97b}.

For practical purpose the vertex functions determined 
as in Ref. \cite{sehn96} are parametrised in the following way 
\beq
 {\overline \Gamma}^{(12)}_{s,0} (\rho_{0}^{(i)},v_{{\rm rel}}) 
 = \alpha_{s,0} e^{-\beta_{s,0}(\rho_{0}^{(1)}+ \rho_{0}^{(2)}) }
 + \gamma_{s,0}
\quad .
\label{gamma_12_fit}
\eeq
This allows an extrapolation to small densities ($\leq 0.5 \rho_{{\rm sat}}$) 
where no DB self-energies have originally been 
available \cite{thm87a} for the construction discussed above. 
In table 2 the coefficients 
$\alpha$, $\beta$ and $ \gamma$, Eq. (\ref{gamma_12_fit}), are given 
for different streaming velocities $v$ in the c.m. system of 
the currents. In symmetric systems these are 
$u^{(1)}_\mu =(\gamma,\gamma v) $ 
and $u^{(2)}_\mu =(\gamma,-\gamma v) $ with the relative velocity 
$v_{{\rm rel}}= \frac{2v}{1+v^2}$. 
In Fig. \ref{gamma_12_graph} the 
${\overline \Gamma}^{(12)}_{s,0}$ are shown in dimensionless units as 
functions of the subsystem densities $\rho_{0}^{(i)}$ and the c.m. streaming 
velocities $v $. The overall behavior of the scalar and 
vector vertex functions is quite similar. 
Both generally decrease with increasing 
density, however, the momentum or velocity dependence is more complex 
\cite{sehn96}. It should be kept in mind  
that the non--relativistic mean field involves a cancellation of the scalar and 
vector self-energies (see Eq. (\ref{opt_LCA}), below) and thus reacts 
sensitively to small variations 
of these functions. The corresponding coupling constants 
of the $\sigma\omega$--model (QHD1) \cite{sw86} 
are $\frac{g_\sigma^2}{m_\sigma^2}\frac{M^2}{4\pi} = 21.25 $ and 
$\frac{g_\omega^2}{m_\omega^2}\frac{M^2}{4\pi} = 15.95$.

The vertex functions, Eq. (\ref{gamma_12_fit}), and the 
self-energy, Eq. (\ref{sigma_12_H}) are now used to approximate the 
non--equilibrium self-energy $\Sigma^{+}_H$ in the kinetic 
equation (\ref{TP5}) in the spirit of the LCA, i.e. by extracting 
locally the parameters $p_{F_i} (x)$ and $v_{{\rm rel}} (x)$. 
In the local density approximation (DB/LDA) these coupling functions 
are used at zero relative velocity and at the total density 
${\overline \Gamma}^{(12)}_{s,0} (\rho=\rho_{0}^{(1)}+\rho_{0}^{(1)},
v_{{\rm rel}}=0)$.

A useful quantity to characterize the mean field is the real part of 
the optical potential. The nucleon--nucleus optical 
potential is well known from proton 
scattering data \cite{hama90} and has been used to fit the momentum 
dependence of mean fields \cite{ai91,wbc92}. 
The quantity of interest in a heavy ion collision is 
the nucleon optical potential in a nucleus--nucleus collision. 
In colliding nuclear matter it is obtained as \cite{sehn96} 
\beq
\label{opt_LCA}
 Re U_{{\rm opt}} (E) = Re {\overline\Sigma}^{(12)}_{s} - 
\left( \frac{E}{M}+1 \right) Re {\overline\Sigma}^{(12)}_{0} 
+ \frac{ ( Re {\overline\Sigma}^{(12)}_{s} )^2 -
         ( Re {\overline\Sigma}^{(12)}_{0} )^2 }{2M}
\eeq 
with $E$ the mean incident energy of a nucleon in the colliding 
system. It depends on the incident energy $E$ also through the relative 
velocity $v_{{\rm rel}}$ in the self-energies in Eq. (\ref{sigma_LCA_1}). 
A corresponding potential in the LDA is the nucleon optical potential of a 
nucleon of energy $E$ in nuclear matter. It is given also by expression 
(\ref{opt_LCA}) whith $\Sigma^{(12)}$ replaced by the 
equilibrium nuclear matter self--energies 
taken at the total density $\rho_{\rm B} =\rho_{0}^{(1)}+ \rho_{0}^{(2)}$. 
Thus, refering back to Fig. \ref{LCA_graph}, the nucleon--nucleus potential 
is experienced by a nucleon in the left configuration, i.e. a distance 
$P_{inc}$ away 
from the center, while the potential in the nucleus--nucleus collision 
is the potential experienced by a nucleon in the mean in the right 
two--ellipsoid configuration.

These optical potentials are compared 
in Fig. \ref{opt_graph}. The DB nucleon--nucleus optical potential 
\cite{thm87a} at saturation density and two times saturation density is 
compared to the optical potential of a mean nucleon in 
a nucleus--nucleus collision. 
As can be seen the overall behavior of the different potentials 
is similar, in that the behavior with increasing energy is dominated by the 
repulsive vector fields. However, the nucleus--nucleus optical potential 
in general is less repulsive. This effect increases in significance with 
increasing density and is quite distinct at 
$2\rho_{{\rm sat}}$, a density 
which is easily reached in a heavy ion collision. On the average the nucleon in 
the two-stream configuration sees lower momentum components than in the 
one--sphere configuration.

Fig. \ref{opt_graph} also shows the experimental data \cite{hama90} which 
should be compared to the nucleon--nucleus potential. It is seen that the DB results 
of ter Haar and Malfliet \cite{thm87a} used in the present work 
are in a reasonable agreement with the data below 1 GeV incident energy. 
We mention that in Ref. \cite{sefu97} very similar results 
have been obtained for the real part of the optical potential 
by relativistic Brueckner calculations using the Bonn potentials. 
Generally, all these calculations start to overshoot the optical 
potential at energies above 1 GeV. This deviation from the 
empirical values may be due to the fact that the 
underlying NN interaction, i.e. the one--boson--exchange potentials, 
\cite{thm87a,bonn89} are fitted to low energy phase shifts 
up to 300 MeV. Furthermore, with increasing energy 
the excitation of baryonic resonances which is not included 
in standard Brueckner calculations, may start to play an 
important role and also 
particle production starts to influence the reaction dynamics. 
The NL2 parameterization 
of the $\sigma\omega$--model considerable underestimates the data in 
the energy range considered.

\section{Numerical realization}
\subsection{The relativistic Landau-Vlasov method}
In this section we discuss questions related to the numerical simulation of 
the relativistic BUU equation. As in other works we adopt a 
test particle method, however, we use covariant extended 
test particles of Gaussian shape in coordinate and momentum 
rather than point--like test particles \cite{bkm93,koli96}. 
This method was called 
'relativistic Landau--Vlasov' method and is extensively discussed 
in Ref. \cite{fu95}. Here we only 
recall the main features of this approach. 

The $8 \cdot(N \cdot A)$-dimensional phase space distribution of 
the testparticles is given as
\beqa     
(aF)(x,p^*) &=& \frac{C}{N} \sum_{i=1}^{A\cdot N }     
             \int \limits_{-\infty}^{\infty} d\tau     
             g \ls x- x_i\ls\tau\rs\rs  \tilde{g}\ls p^* - p_{i}^{*} \ls\tau\rs\rs    
\nonumber \\     
         &=& \frac{C}{N \ls \pi\sigma\sigma_{p}\rs^3}     
             \sum_{i=1}^{A\cdot N }     
             \int \limits_{-\infty}^{\infty} d\tau     
               e^{(x - x_i (\tau) )^2 /\sigma^2}    
               e^{\ls p^* - p_{i}^{*} \ls \tau\rs\rs^2 / \sigma_{p}^{2}}     
\nonumber \\    
  && \times \delta \left[\ls x_\mu - x_{i\mu} \ls\tau\rs\rs u_{i}^{\mu}     
                 \ls \tau\rs  \right]    
   \delta \left[ p_{\mu}^{*} p_{i}^{*\mu}\ls\tau\rs - m_{i}^{*2} \right]     
\quad .    
\label{fs_1}    
\eeqa
Here $N$ is the number of test particles per baryon and the factor 
$C= \frac{1}{4}(2\pi)^4$ normalizes the phase 
space. $g$ and $\tilde{g}$ are the gaussians shapes of the test particles 
in coordinate and momentum space, and $(x_{i},p_{i})$ are the centers of 
the test particles. 
The $\delta$--function constraints in Eq. (\ref{fs_1}) serve 
to fix the eigentime $\tau$ on the word line, to reduce the phase 
space to $7 \cdot(N \cdot A)$ dimensions, and to 
ensure the arguments of the exponentials to be purely space-like. 
The centers of the gaussians are put on the mass--shell 
$ p_{i}^{*2}=m_{i}^{*2}$. In general the momentum space 
gaussians lead to off--shell contributions and thus 
include a model for the spectral function in a simple way. 
As shown in Ref. \cite{fu95} the 
width of the momentum space gaussian is related to the spectral 
width and hence should evolve dynamically. 
In the present calculations these widths are, however, kept constant. 
For this work the advantage of the use of gaussians lies in a smoother 
representation of the phase space. 

Performing a $p_0$--integration the seven-dimensional distribution 
is obtained as 
\beqa
F(x,{\bf p}^*) &=& \int \frac{d p_{0}}{2\pi} (aF)(x,p^*) 
\nonumber \\
               &=& \frac{C}{N 2\pi} \sum_{i=1}^{A\cdot N }  
               \int \limits_{-\infty}^{\infty} d\tau     
               g \ls x- x_i\ls\tau\rs\rs  \frac{1}{p_{i0}^*} 
               e^{\ls \frac{{\bf p}^{*2}_i}{ p^{*2}_{i0}} 
                      - ({\bf p}^{*} -{\bf p}^{*}_i )^2 \rs / \sigma_{p}^{2}}  
\quad .
\label{fs_1b}    
\eeqa
We also note that this ansatz is consistent with the 
sum rule for the spectral function \cite{btm90}. To evaluate the gaussians 
in Minkowski space the $\tau$--integration over the world line of the particle is 
carried out by expanding the trajectory locally into a Taylor series 
up to first order around an eigentime $\hat \tau$ which 
is fixed by the condition $x_0 = x_{i0}(\htau)$. Then the 
gaussian in Minkowski space reduces to 
\beq 
\int \limits_{-\infty}^{\infty} d\tau     
               g\ls x - x_i \ls\tau\rs \rs =
        (\sqrt{\pi}\sigma)^{-3} e^{R_{i\mu}(x)R_{i}^{\mu}(x) /\sigma^2}  
\eeq
with 
$R_{i}^{\mu}(x) = \ls x^\mu - x_{i}^{\mu} \ls\htau\rs \rs 
                  - \ls x_\nu - x_{i\nu} \ls\htau\rs\rs  
                     u_{i}^{\nu}\ls\htau\rs u_{i}^{\mu}\ls\htau\rs$ 
which is the distance in Minkowski space perpendicular to the 
four--velocity $u_{i} \ls\htau\rs$ of the particle. 

The phase space distribution, Eq. (\ref{fs_1}), provides a 
solution of the Vlasov equation, i.e. the drift 
term of the kinetic equation 
(\ref{TP5}), if the testparticles obey classical 
equations of motion \cite{fu95}
\beqa    
\frac{d x_{i}^\mu  }{d\tau} 
&=& \frac{p_{i}^{*\mu}\ls\tau\rs}{m_{i}^{*}(x_i)}    
\nonumber \\    
\frac{d p_{i}^{*\mu} }{d\tau} 
&=& \frac{p_{i\nu}^{*} \ls\tau\rs}{m_{i}^{*}(x_i)}    
 F^{\mu\nu} \ls x_i\ls\tau\rs\rs + \p^\mu m_{i}^{*}(x_i)    
\quad .    
\label{bgl}    
\eeqa
The DB mean fields in LCA and LDA are determined in every time step as described in 
the previous section. 
For the simulation of the collision term we adopt the full 
ensemble method \cite{bg88} and use the Cugnon parametrisation 
of the NN cross section \cite{cug81}. In the considered energy 
range inelastic channels already play an important role \cite{koli96}, in 
particular the $\Delta$(1235)--resonance influences the dynamics 
to a large extent. The production 
and the decay of $\Delta$--resonances as well as pion production and 
reabsorption are included as described in Ref. \cite{wolf90}. 
The initialization of the nuclei is performed as described in Ref. 
\cite{fu95}.  
\subsection{Fragment content in RLV}
The RLV method is a successful tool to 
simulate heavy ion reactions in the framework of relativistic 
transport theory. It accurately reproduces  
the one-body dynamics of the colliding system \cite{koli96} 
and thus yields reliable results of global flow observables. 
This will be discussed in section 6. 
One--body models, however, do not contain dynamical fluctuations which become 
significant in instability situations in the decompression 
stage of the reaction, and which are important for the 
formation of fragments. The theoretical description 
of higher order correlations and hence of dynamical fragment formation 
is actively debated, and there have been various attempts to describe the 
higher order correlations approximately: by adding a 
fluctuation term leading to 
the Boltzmann Langevin equation \cite{ay90}, 
by choosing the numerical fluctuations 
in a judicious way by the number of test particles \cite{co93}, 
or by introducing 
fluctuations directly into the phase space distribution \cite{co96}. 

In this work we will not address this question. However,  
we need information about the fragmentation in 
the final state of the reaction in order to 
perform a reliable comparison with 
experiments which are sensitive to the fragments via detector acceptances. 
Therefore, we will restrict 
ourselves to a simple model which 
generates intermediate mass fragments. We then 
adopt the FOPI filter simulation 
(see next section) which depends on the fragment charge.

A simple way to generate fragments from the final phase space 
is the phase space coalescence model \cite{kap80,saya81}. It is a method to 
generate many--body correlations which are consistent with a given 
one--body distribution function. Within this model a number of 
nucleons form a fragment if their distances 
in coordinate as well as in momentum 
space are smaller than certain coalescence parameters $R_{c}$ 
and $P_{c}$, respectively 
\begin{displaymath}
| \vec{x}_{i}-\vec{X}_f | \leq R_c \qquad\mbox{,} \qquad 
| \vec{p}_{i}-\vec{P}_f | \leq P_c
\qquad .
\end{displaymath}
Here $\vec{x}_{i}$ and $\vec{p}_{i}$ are the coordinates of the $i$-th 
nucleon in coordinate and momentum space, respectively, 
and $\vec{X}_{f}$ and $\vec{P}_{f}$ are the center-of-mass 
coordinates of the fragment. 
From the distribution of the $N \cdot A$ test particles of a simulation, 
$A$ test particles 
are chosen at random and the above procedure is applied at the final time step 
of the calculation. This generates a fragmentation "event", and to generate 
distributions this procedure is repeated many times ($\approx 10000$). 
Thus the fragment formation 
is modeled by the choice of the parameters $R_{c}$ and $P_{c}$ 
which are effective parameters. They are determined by a 
fit to the experimental charge distributions. 
This procedure has to be carried out for each interaction separately. 

In order to give an impression of this procedure we show in 
Fig. \ref{fig4.2} the charge distributions obtained for 
two different mean fields,  the non-equilibrium DB forces (DB/LCA) 
and the non--linear $\sigma\omega$ model NL2. As an example we use central and 
peripheral Au on Au reactions at different incident energies. 
The experimental curves are taken from FOPI 
\cite{reis97} and represent $4 \pi$ 
charge distributions extrapolated within the blast model \cite{bon78}. 
The coalescence parameters have been adjusted such that both theoretical 
distributions yield similar abundances of light 
fragments ($Z \leq 3$). However, the DB model in LCA (and in LDA not 
shown here) produces too many heavy fragments ($Z \geq 3$). 
This is also evident from test particle distributions (not shown here), 
which are considerably more "lumpy" that for 
NL2. The reason may be due to the extrapolation 
of the DB nuclear matter results to densities below half saturation density 
(see the discussion in section 4). In any case, deviations for higher $Z$ between 
the experimental and theoretical distributions will have no consequence in the 
following flow analysis, since the absolute 
yields of heavy fragments ($Z \geq 3$) is exponentially decreasing and 
multiplicity distributions and flow observables are governed by 
the dynamics of protons and light fragments. The fit procedure, which 
has been performed in the energy range from $250$ to $800$ A.MeV, 
ensures a correct inclusion of detector acceptance cuts simulated by the 
filter program. The coalescence parameters $R_{c}\sim 4 ~fm $ and 
$P_{c}\simeq 1.4 ~fm^{-1}$ are similar to values obtained 
from QMD simulations \cite{ai91}.

\section{Flow analysis}

We now study the global flow observables and compare them to 
experiments, in particular we discuss recent flow data measured by the 
FOPI collaboration 
\cite{bastid97,croch1,crochet97,croch2,crochthesis,croch3}. These 
flow data were obtained with the Phase I of the FOPI detector at 
the SIS/ESR accelerator facility at the GSI. 
For details we refer to work of the FOPI 
collaboration \cite{fopi1,bastid97,croch1} and references therein. 
Relative to earlier experimental investigations \cite{flow1} these new 
experiments have the advantage of high selectivity concerning the 
detection of intermediate mass ($Z \geq 3$) fragments up to $Z=15$. 
This ability of the FOPI detector allows a cleaner 
identification of the collective flow 
signal and thus provides flow data of high precision for different fragments 
over an energy range of several hundred A.MeV. A wide range of the impact 
parameters has been explored which offers the possibility 
to study the centrality 
dependence of the observables. We discuss flow observables 
concerning the reaction dynamics  
in the reaction plane (in--plane flow) and perpendicular 
to the reaction plane (out--of--plane flow).

A reliable comparison to the data requires to account for the 
detector acceptances, such as geometrical 
cuts and thresholds, which is achieved by the FOPI filter simulation 
code \cite{reis}. Fragments are generated as outlined above. 
Furthermore, the correspondence between the centrality classes of the reaction 
(impact parameters) and the multiplicities of the detected charged 
particles (PM) (see next section) has to be determined. 
In the following analysis we choose the $z$--axis as 
the beam direction, and the 
reaction plane as the $xz$--plane, where the projectile 
and target are asymptotically shifted by the impact parameter $b$ in the 
$x$--direction. 

\subsection{Centrality selection}

A problem in any experimental analysis is that it does not allow a direct 
determination of the impact parameter of the reaction. 
The usual procedure is to 
extract the centrality of the reaction from the measurement of 
strongly impact parameter dependent observables. The correlation 
between these quantities and an impact parameter range or centrality 
class has to be obtained in a model. In the FOPI 
experiments several observables 
have been investigated to select the centrality class of the reaction 
\cite{reis97,fopi1,bastid97}. One is the $ERAT$-ratio  
\begin{equation}
ERAT = \Bigg( \sum_{i=1} \frac{ E^{(i)}_{\perp} }
{ E^{(i)}_{\|} } \Bigg)_{Y^{(i)} \geq Y_{{\rm c.m.}}}
\label{erat}
\end{equation}
defined by the ratio of the transverse $E_{\perp}$   
and the longitudinal kinetic energy $E_{\|}$. 
The sum in Eq. (\ref{erat}) runs 
over all nucleons in the forward hemisphere, characterized by 
$Y^{(i)} \geq Y_{{\rm c.m.}}$ ($Y^{(i)}$ being the rapidity of particle $i$). 
Thus the $ERAT$ value is a measure of the amount 
of the kinetic energy transferred perpendicular to the beam axis. 
The pronounced impact parameter dependence is illustrated in 
Fig. \ref{fig5.1a} where the $ERAT$ ratio at different energies is shown 
as a function of the impact parameter. Large (small) 
values of $ERAT$ correspond to central (peripheral) reactions. 
Due to the strong variation at small values of $b$ the $ERAT$ observable 
is particularly suited for the selection of very central reactions 
\cite{reis97}. In our previous comparison \cite{fgw96} to flow data 
\cite{fopi1} the $ERAT$ distributions has been used to select the 
centrality classes of the reactions. A detailed 
discussion of the $ERAT$ selection can be found in the work of 
Reisdorf et al. \cite{reis97}. 

Another suitable observable to determine the centrality class 
of the reaction is the charged particle multiplicity $PM$. 
This observable measures the number of detected particles 
with charge $Z \geq 1$ 
from the participant matter. Thus, high (low) multiplicity 
values correspond to 
central (peripheral) reactions. Such multiplicity distributions 
show a typical plateau for small $PM$--values and a 
strong decrease with increasing multiplicity \cite{reis97}. 
Thus they are sensitive to peripheral collisions. The multiplicity 
bins which correspond to different 
centrality classes, are usually defined in the following way: 
the lower limit of the highest multiplicity 
bin $PM5$ is fixed at half of the plateau value, and the 
remaining multiplicity 
range is divided into four equally spaced intervals, 
denoted by $PM4$ to $PM1$. 
$PM5$ then corresponds to most central reactions 
\cite{reis97,bastid97,croch1,crochet97,croch2}.

The theoretical charge distributions 
(see Fig. \ref{fig4.2}) and the corresponding multiplicity distributions are 
model dependent. Hence, the correlation between the charged-particle 
multiplicity $PM$ and the centrality class has to be established separately for 
each model and for different energies. 
Table \ref{pm-tab} gives  for the different models 
the lower limits $PM_{L}$ for the highest 
multiplicity bin $PM5$. 
The corresponding experimental values are taken from Ref. \cite{croch1}. 
The theoretical determination of the impact parameter 
range $[b,b+\Delta b]$ follows from the calibration curves $PM$ versus $b$. 
Fig. \ref{fig5.1b} shows this correlation between the $PM$ multiplicity 
and the impact parameter $b$ for two of the mean fields applied. 
Both curves show the typical decrease of $PM$ with increasing 
impact parameter. However, the absolute $PM$ values differ due to 
different charge distributions (see Fig. \ref{fig4.2}). 
As discussed above, the DB/LCA approach, and very similarly also 
DB/LDA (not shown here), yield more heavy clusters 
and therefore a smaller total yield of charged particles
than NL2. The centrality calibration used here 
takes such facts into account and provides the 
correct correlation between the impact parameter range and the 
multiplicity bins for each model. In table \ref{bmean-tab} mean impact parameters 
\begin{displaymath}
< b > = \frac{ \int_{b_{{\rm min}}}^{b_{{\rm max}}} b \cdot b db }
{ \int_{b_{{\rm min}}}^{b_{{\rm max}}} b db} = \frac{2}{3} 
\frac{ ( b_{{\rm max}}^3-b_{{\rm min}}^3 )}
{( b_{{\rm max}}^2-b_{{\rm min}}^2 )} 
\end{displaymath}
are given for the system Au on Au at the various energies considered. 

\subsection{In-plane flow}

An important observable to characterize the dynamical 
evolution of the reaction is the mean transverse momentum 
projected on the reaction plane $< p_{x}(Y_{{\rm c.m.}}) >$ as a function 
of the center--of--mass rapidity $Y_{{\rm c.m.}}$, often also 
denoted as "bounce-off", 
"transverse flow" or "sideward flow". It is sensitive essentially only to 
the mean field, in particular on its repulsiveness, and has  
thus been regarded as a source of information on the nuclear EOS 
\cite{bkm93,pei89}. For a first discussion  
we compare in Fig. \ref{fig5.2a} the sideward flow per 
nucleon as a function of 
the normalized rapidity $Y^{(0)} = Y_{{\rm c.m.}}/ Y_{{\rm proj}}$ 
obtained with the DB/LCA and NL2 models in a semi--central 
Au on Au reaction at $600$ A.MeV. 
In a comparison to experiment one has to take into account  
the influence of the acceptance filter on the in--plane flow. In 
previous works \cite{reis97,fopi1,bastid97,croch1,crochet97} 
mainly in the non-relativistic QMD model it was found that the 
FOPI filter scarcely affects the shape of the in--plane flow 
in the forward hemisphere ($Y^{(0)} \geq 0$). This is confirmed in the 
present calculations, Fig. \ref{fig5.2a}, 
where the filtered as well as the unfiltered in-plane flow 
is shown. Therefore this observable can reliably be compared to the data.

The results of Fig. \ref{fig5.2a} also show the 
strongly repulsive character of the DB/LCA mean fields relative to NL2. 
We note that the 
repulsion in relativistic transport models originates mainly from the 
momentum dependence, in contrast to non-relativistic approaches as 
BUU \cite{bg88} or QMD \cite{ai91} where the density dependence is decisive. 
Although both models, DB and NL2, yield a similar density dependence, i.e. 
comparable slopes of the EOS at high densities, as seen in Fig. \ref{EOS_graph}, 
the reaction dynamics is very different in the two cases. Thus the 
sideward flow generated by the repulsive momentum 
dependent part of the field is different in the two models as already 
seen in the optical potentials of Fig. \ref{opt_graph}.  
A scaling behavior of the in--plane flow with the inverse of the value of 
$m^{\ast}$ has, e.g., been discussed in Ref. \cite{fu96b}.

We have analyzed the sideward flow with respect to centrality and fragment 
charge, as in the experimental data. We start the discussion with the 
sideward flow of protons ($Z=1$) 
shown in Fig. \ref{fig5.2b}. The theoretical results are compared 
to the FOPI data for the system Au on Au 
at incident energies of 250, 400 and 600 A.MeV 
and for two typical centrality classes, i.e.  
$PM5$ (central collisions) and $PM4$ (semi-central reactions). 
For the relation between the 
impact parameter ranges and the multiplicity bins we refer to 
table \ref{bmean-tab}. We find that the DB/LCA mean 
fields are generally able to  
reproduce the sideward flow of protons over the considered energy 
range. In particular, for the most central 
($PM5$) reactions the agreement between the DB/LCA results 
and the data is excellent for the range 
$0 \stackrel{\le}{~} Y^{(0)} \stackrel{\le}{~} 1.2$. 
At higher rapidities slight deviations occur which, 
however, could also be due to statistical fluctuations. 
Comparing to the local density approximation DB/LDA we see that 
non-equilibrium effects are most pronounced in central collisions (PM5). 
Here DB/LDA yields generally a larger flow and strongly 
overestimates the data. With increasing impact parameter the description 
with DB/LCA becomes less accurate and slightly overestimates the data. 
This behavior was already observed in 
Ref. \cite{fgw96}. This may be due to a simplification in our treatment in  
that the non--equilibrium mean fields 
are always determined for symmetric configurations \cite{sehn96}, 
i.e. the local $2$--Fermi--sphere momentum configurations are symmetrized 
at $p_{F_1}=p_{F_2}$. Such 
a treatment should be better in central collisions which exhibit a 
high symmetry of the local phase space. In more peripheral collisions, 
however, asymmetric configurations where the densities of projectile 
and target are significantly different become more important and the 
approximation is less accurate. 

The $\sigma\omega$ mean field (NL2), on the other hand, leads to substantial 
underestimation of the in--plane flow for all incident energies and 
centrality classes. Obviously the momentum dependent repulsion is too 
small in this model.

In Figs. \ref{fig5.2bz2}, \ref{fig5.2bz3} and \ref{fig5.2bz4} 
the in-plane flow per nucleon $<p_x / A> $ 
for light fragments with charges $Z=2,3$ and $4$, respectively, is 
shown at incident energies of $250$, $400$ and $600$ A.MeV 
and for the same centrality classes as in Fig. \ref{fig5.2b}. First 
of all, it is seen that all models are able to qualitatively reproduce 
the increasing in--plane flow per nucleon with increasing fragments size. 
The enhanced flow most likely results 
from the fact that these large fragments are created in the cool spectator 
regions. In 
Ref. \cite{essler} it was also found that thermodynamical 
instabilities appear after the compression phase in the 
cool spectator regions. Such a scenario is also supported 
by QMD calculations where it is found that heavy fragments 
show a smaller stopping than protons \cite{goss97}. The dependence of the 
fragment flow on the mean field model is again pronounced. 
As in the case of protons the DB model 
generally yields a higher in-plane flow  than NL2. 
The importance of non-equilibrium effects seems 
to decrease with increasing fragment size. 
E.g. for Z=4 there are no significant differences visible 
between DB/LCA and DB/LDA whereas for Z=2 the situation is still 
similar to the case of Z=1. However, the comparison of the different calculations 
with the data is not so conclusive. For peripheral reactions ($PM4$) the 
DB/LCA approach is able to reproduce better the fragment dynamics 
whereas NL2 again underestimates the data for rapidities 
greater than $\sim 0.5$. However, for central 
reactions ($PM5$) the opposite trend is observed, 
i.e. the DB calculations overestimate the data whereas NL2 
only slightly underestimates them. Thus, a reliable comparison to 
dynamical observables derived from fragment distributions 
seems problematic. Theoretical predictions could be strongly 
influenced by the models of fragment formation and 
a simple phase space coalescence model could 
be insufficient in order to draw reliable conclusions. 
Such problems are perhaps smaller when integrated observables are considered,  
which we do now.

The mean directed in-plane flow per particle 
$P_{x}^{{\mathrm dir}}/A$ in one event is defined as 
\begin{equation}
P_{x}^{{\mathrm dir}}/A = \frac{ \sum_{i=1}^{M} p_{x}^{(i)} 
{\rm sign} (Y^{(0)}_{i}) }{ \sum_{i=1}^M A_i }
\label{pxdir}
\end{equation}
where the sum in Eq. (\ref{pxdir}) runs over all $M$ fragments    
of an event in a given centrality class and $A_i$ and $p_{x}^{(i)}$ is the mass number 
and transverse momentum of the $i$--th particle, respectively. 
$P_{x}^{{\mathrm dir}}$ is a very useful observable 
to classify the global reaction dynamics, because, except for detector cuts, 
it is independent of the fragment generation procedure. 
Fig. \ref{fig5.2c} shows the energy dependence of $<P_{x}^{{\mathrm dir}}/A>$, i.e. 
averaged over many events. It is seen that the non-equilibrium 
DB mean fields are able to reproduce the correct energy dependence of the 
directed in-plane flow over the energy range from $250$ up 
to $600$ A.MeV. At $800$ A.MeV it slightly overestimates the data, 
however, at this energy the DB approach \cite{thm87a,sefu97} also 
starts to overestimate the empirical optical potential (real part). 
Neglecting non-equilibrium effects (DB/LDA) generally leads 
to a larger flow. The relative importance of these effects is 
small at 250 A.MeV, maximal at 400 A.MeV and then decreases again with 
increasing energy. This energy dependence is reasonable since at 
250 A.MeV the initial relative velocity of the local nuclear matter 
currents, i.e. the anisotropy of the phase space, is still small 
and at high energies elastic and inelastic NN--scattering processes 
start to dominate the reaction dynamics. Interestingly, the mean field 
effects are most pronounced at 400 A.MeV where the sideward flow 
excitation function, i.e. the flow scaled by the beam energy, is 
maximal as well \cite{ritter97}.

Fig. \ref{fig5.2d} illustrates the centrality dependence of 
$P_{x}^{{\mathrm dir}}$ at 400 and 600 A.MeV. The dependence of 
$P_{x}^{{\mathrm dir}}$ on the impact parameter shown 
in the left panels does not depend strongly on the   
energy in the two cases. The in-plane flow is 
maximal at impact parameters between 4--5 fm. It rapidly drops down 
to zero with decreasing impact parameter which reflects the symmetry 
around the beam axis in very central collisions. The DB/LCA mean field 
yields a maximum value of $P_{x}^{{\mathrm dir}}$ which is about twice 
as large as that reached with the softer NL2 model, which was already seen in 
Fig. \ref{fig5.2b}. Applying the local density approximation DB/LDA 
the flow signal is again increased by about 30\% compared to DB/LCA. 
As already seen in the previous figure this effect is maximal at 
400 A.MeV. In the right panels of 
Fig. \ref{fig5.2d} the directed flow as a function of the centrality class 
determined with the multiplicity selection is 
compared to the data \cite{crochthesis}. The $PM$ dependence of   
$P_{x}^{{\mathrm dir}}$ is obtained by averaging 
$P_{x}^{{\mathrm dir}} (b) $ over the corresponding $PM$ 
classes given in Tab. \ref{bmean-tab}  
\begin{equation}
<P_{x}^{{\mathrm dir}}(PM)> = \frac{ \int_{b_{{\rm min}}}^{b_{{\rm max}}} 
P_{x}^{{\mathrm dir}} b^{2} db }
{ \int_{b_{{\rm min}}}^{b_{{\rm max}}} b^{2} db }
\quad .
\end{equation}
In this context it should be noticed 
that the $PM$ selection is less accurate for 
very central collisions. At small impact parameters the multiplicities almost 
saturate, see Fig. \ref{fig5.1b}, which leads to an insufficient 
$b$ resolution below $b \leq  2 - 3$ fm.  Consequently, the measured 
$P_{x}^{{\mathrm dir}} (PM)$ is actually not zero for the highest 
multiplicity, but corresponds to central reactions at $b\sim 3$ fm 
\cite{reis97,croch1,croch2}. A better 
resolution of very central events can be achieved by the ERAT selection, 
see Fig. \ref{fig5.1a}. This problem was extensively discussed 
in Ref. \cite{reis97}. 
The comparison to the data in Fig. \ref{fig5.2d} clearly shows that 
the DB/LCA calculations are able to reproduce the in-plane flow for central 
up to semi-central reactions. Here the local density approximation 
DB/LDA again leads to a strong overestimation of the data. Only 
in peripheral collisions corresponding to 
low multiplicities, $ PM/PM5_L \sim 0.4$, the 
local density approximation is reliable.  

\subsection{Out-of-plane flow}

The emission of nuclear matter perpendicular 
to the reaction plane, the so--called "squeeze-out", is another 
characteristic feature of heavy ion collisions. This 
effect is mainly caused by the fireball expansion and 
the shadowing by spectator matter in the reaction plane. 
The out-of-plane emission is therefore most 
pronounced close to the mid-rapidity region. It was predicted by early 
hydrodynamical calculations \cite{kapusta81,stoecker82} and 
later confirmed experimentally \cite{gut89a,diogene90}. Since the 
out-of-plane emission mainly originates from the  
participant region it contains direct information on the highly  
compressed nuclear matter and is sensitive on 
the nuclear EOS \cite{har94,bass95}.

Before analyzing the squeeze--out in detail 
it is helpful to investigate its formation, i.e. the 
transition from the in-plane to the out-of-plane flow. Azimuthal distributions 
around the beam direction $dN/d\Phi$ (where $\Phi$ is the azimuth of the 
particles with respect to the reaction plane) with changing  
center-of-mass polar angle $\Theta_{{\rm c.m.}}$ 
are suitable to study the formation of squeeze-out \cite{crochet97}. 
Fig. \ref{fig5.3a} 
compares azimuthal emission patterns at different polar angles 
for a semi--central ($PM4$) Au on Au reaction at $600$ A.MeV to the 
FOPI data \cite{crochet97}. 
At forward polar angles the distributions exhibit strongly 
enhanced in-plane emission along the sideward flow direction of the 
projectile ($\Phi=0^{o}$ and $360^{o}$). Increasing $\Theta_{{\rm c.m.}}$ a sudden  
change in the azimuthal emission pattern is observed for 
$\Theta_{{\rm c.m.}} \stackrel{\ge}{~} 70^{o}$. 
At $\Theta_{{\rm c.m.}} \approx 90^{o}$ 
the azimuthal distributions clearly show maxima at $\Phi=90^{o}$ 
and $\Phi=270^{o}$ which are evidence for the squeeze--out. Both models, 
DB/LCA and NL2, are in qualitative agreement with 
the data, for NL2 it is almost quantitative. This indicates that the 
dynamical evolution of the 
out--of--plane emission is correctly described by 
these forces. On the other hand, DB/LDA leads to an extremely 
strong squeeze--out signal for large polar angles approaching 
$\Theta_{{\rm c.m.}} = 90^{o}$. The particles around 
$\Theta_{{\rm c.m.}} = 90^{o}$ preferentially belong to the 
stopped fireball matter and the strong repulsion of the 
DB/LDA force seems to push out these particles too rapidly 
from the fireball region resulting in a too large squeeze--out. 

As a quantitative measure of this transition from in--plane to out--of--plane flow
the anisotropy ratio $R$ has been proposed \cite{crochet97}
\begin{equation}
R = \frac{ { dN/d\Phi \mid }_{ 0^{o} < \Phi < 45^{o} } + 
           { dN/d\Phi \mid }_{ 315^{o} < \Phi < 360^{o} } }
         { { dN/d\Phi \mid }_{ 135^{o} < \Phi < 225^{o} } } \quad ,
\label{ratio}
\end{equation}
considered as a function of $\Theta_{{\rm c.m.}}$. It was found that 
at intermediate polar angles the FOPI filter scarcely affects 
the anisotropy ratio $R$ and a reliable comparison to the data 
\cite{crochet97} can be performed. This is done in the Fig. 
\ref{fig5.3c}. It is seen that the DB mean fields yield 
a significantly higher anisotropy than the 
NL2 model and strongly overestimate the data at 
intermediate and larger polar angles. Since the squeeze--out 
effect which is maximal at $\Phi= 90^{o}, 270^{o}$ degree, is not 
covered by the ratio defined in Eq. (\ref{ratio}) DB/LCA and 
DB/LDA yields similar results for $R$ although they have 
completely different out--of--plane emission patterns. 
The NL2 model leads 
to a much smoother $\Theta_{{\rm c.m.}}$--dependence of $R$ 
but generally underestimates the data. 
Comparing with QMD calculations \cite{crochet97} 
it seems that the momentum dependence of the 
DB mean forces is responsible for the overestimation of $R$. Momentum 
dependent Skyrme forces (hard and soft) yield qualitatively 
similar results as DB whereas a static 
hard Skyrme force gives results closer to the data and more 
comparable to NL2. On the other hand, the momentum dependence in 
the last section was found to be essential for the 
correct in-plane dynamics which is reflected in Fig. \ref{fig5.3c} by the 
observation that the DB/LCA calculation is closer 
to the data at forward angles. This indicates that the strong 
repulsion provided by the DB model yields a reasonable description of 
the in--plane dynamics of the spectator matter but the model has 
problems to describe exactly the flow evolution of the stopped matter in the 
fireball region. Thus, additional effects as in-medium 
modifications of the NN cross section \cite{cross,thm87b,lima93,jae92} may 
play a role. However, non-equilibrium effects (DB/LCA) again 
turn out to be essential and improve the agreement with the 
experimental observations.

We now discuss the out-of-plane emission of participant matter 
at mid-rapidity, i.e. 
$-0.15 \le Y^{(0)} \le 0.15$ where the squeeze-out signal is maximal 
\cite{bastid97,croch1,croch2,croch3,crochthesis}.  
To take the limited detector acceptance into account means  
to make cuts with respect to the normalized transverse momentum per nucleon 
$P_{T}^{(0)} = \frac{ (P_{T}/A) }{ (P_{T}^{{\rm proj}}/A_p) }$ 
\cite{bastid97,croch1,croch2}. Fig. \ref{fig5.3e} shows 
the influence of the FOPI 
filter on the squeeze--out, for NL2 as an example. 
The calculations are performed with and without  
the FOPI filter and a $P_{T}^{(0)}$--cut 
($0.06 \le P_{T}^{(0)} \le 0.55$), respectively. 
The solid lines are fits to the theoretical distributions according to  
\begin{equation}
N(\phi) = \alpha_{0} \left( 1 + \alpha_{1}\cos(\phi) 
+ \alpha_{2}\cos(2\phi) \right)
\label{fit}
\qquad .
\label{squeeze1}
\end{equation}

From Eq. (\ref{squeeze1}) a quantity called the 
squeeze--out ratio is obtained as 
\beq
R_{N} = \frac{ 1-\alpha_{2} }{ 1+\alpha_{2} }
\quad .
\label{sqratio}
\eeq 
If the $P_{T}^{(0)}$ cuts are taken into account 
the filter has only a small influence within the 
statistical uncertainties 
($R_{N} = 1.226 \pm 0.04$ without filter and cuts and 
$R_{N} = 1.15 \pm 0.03$, $1.168 \pm 0.03$ with $P_{T}^{(0)}$--cut but 
without and with filter, respectively). A similar observation was 
made in Ref. \cite{bastid97}.

In Fig. \ref{fig5.3f} the model dependence of the squeeze--out signal 
is considered, i.e. the 
results for the DB forces DB/LCA and DB/LDA and 
the NL2 are compared. The calculations are performed 
for a semi-central (PM4) Au on Au reaction at $600$ A.MeV and 
the FOPI filter has been applied. As already observed in Fig. 
\ref{fig5.3c} we see a strong dependence of the out--of--plane 
flow on non-equilibrium effects. The local density approach 
to the DB mean field (DB/LDA) leads to a much stronger out--of--plane 
flow compared to the case when non-equilibrium effects are taken into 
account (DB/LCA). On the other hand, DB/LCA still gives a stronger 
signal than the softer NL2 model. 
Thus the out--of--plane flow shows the same general dependence on the 
nuclear mean field as the in--plane--flow. 
In the upper panel of Fig. \ref{fig5.3f} we compare the NL2 result 
to the corresponding FOPI data \cite{croch2} for both, $Z=1$ and $Z=2$.
The theoretical curve somewhat overestimated the 
squeeze--out signal for $Z=1$ but is in agreement with the $Z=2$ data. 
Since the experimental results are partially biased by detector 
inefficiencies \cite{croch2} but exhibit a strong charge 
dependence it seems to be more reasonable to compare in average 
to the $Z=1$ and $Z=2$ data. Doing this, NL2 and also DB/LCA 
slightly overestimate the data but are in qualitative agreement 
whereas DB/LDA strongly overpredicts the out--of--plane emission.

Selecting particles with high transverse momentum an even 
more pronounced dependence on the EOS has been observed 
\cite{croch2}. Thus, in the following we concentrate on the analysis 
of high $P_{T}^{(0)}$--particles, which are selected by a 
cut $0.5 \leq  P_{T}^{(0)} \leq 0.55$. 
Fig. \ref{fig5.3g} shows the corresponding $R_{N}$ ratio Eq. (\ref{sqratio}) 
in semi--central reactions ($PM4$) as a function of the incident energy. 
Here the results for DB/LDA are not shown since they have already 
been shown to be completely off the data. For the high energetic particles 
NL2 again underestimates whereas DB/LCA tends 
to overestimate the data. Only at the highest energy considered, i.e. 
at 800 A.MeV, the calculations agree with the experiment. 
Except for the lowest data point both models miss the experimental curve 
equally, however, due to the narrow energy window the 
statistical errors of the 
calculations are rather large, in particular at 250 A.MeV. 

Both models, DB and NL2, do not provide 
a quantitative description of the out--of--plane dynamics. 
The DB forces generally yield larger squeeze--out 
and anisotropy ratios which is consistent with the in--plane analysis 
and can be explained with the repulsive character of the nuclear 
fields. It is clear, however, that the inclusion of non-equilibrium effects 
is essential for a reasonable description of the reaction dynamics. 
In-medium modifications of the NN cross section 
may further influence the azimuthal distributions since the shadowing 
effects by the spectator matter are strongly governed by binary collisions. 


\section{Summary and conclusions}

We studied the reaction dynamics in intermediate energy 
heavy ion collisions in the framework of relativistic transport 
theory and compared collective in--plane and 
out--of--plane flow observables to 
experimental data obtained by the FOPI Collaboration. 
In order to obtain a more complete 
picture these investigations covered a wide 
range of incident energies ($250$ to $800$ A.MeV) and 
the full centrality range ($0 \leq b [fm] \leq 14$). 
The limitations of detector acceptances were 
taken into account by generating fragment 
distributions by a coalescence model. 

The nuclear mean field used in the transport calculations 
was based on relativistic Dirac-Brueckner (DB) theory. 
We argue that the colliding system is far away from global and even local 
equilibrium during most of the reaction and therefore a local 
density approximation (LDA) of 
the mean field may not be justified. Thus we account 
for the non--equilibrium features of the dynamical phase 
space configuration also on the level of the mean field. 
The mean field is parameterized for a class of 
anisotropic phase space configurations given by $2$--Fermi--ellipsoid 
configurations in momentum space. We called this approximation 
" local (phase space) configuration approximation (LCA)" in contrast to 
the commonly used local density approximation (LDA). 
We further compared this microscopic 
approach to a non--linear parameterization of the $\sigma\omega$ model 
(NL2) which provides similar density dependence, i.e. 
a similar nuclear matter equation-of-state, as the DB model, 
however, a significantly less repulsive optical potential.   

The analysis of in--plane flow observables 
demonstrates that the DB mean fields 
are generally able to provide a reliable description of the reaction dynamics. 
However, the inclusion of non-equilibrium features in the interaction 
is of essential importance, otherwise the in--plane flow is 
mostly overpredicted. The use of Dirac--Brueckner 
fields is, however, only reliable below about 
1 GeV/nucleon as seen from a comparison to the 
empirical nucleon--nucleus optical potential. This limitation is 
reflected in the analysis 
of in--plane flow observables where the agreement with the data 
is best below 800 A.MeV. 
The calculations with the more weakly repulsive non--linear 
$\sigma\omega$ model (NL2) generally underestimate  
the in--plane flow. This demonstrates that the reaction dynamics 
is governed mainly by the repulsive momentum dependence of 
the nuclear fields and less so by the density dependence in the 
compression phase. 
The in--plane flow of light fragments further opens 
the possibility to test the model at low densities 
since the fragment formation mainly 
takes place in the decompression phase. However, in this study  
the flow of fragments does not lead to decisive results. 
One reason may be the simple theoretical 
description of the fragment formation. On the other hand, the 
DB forces used in the present calculations were extrapolated into 
the low density regime and thus are not very reliable there.  
Future investigations of the fragment flow might 
be able to set constraints on the low density behavior of the 
nuclear mean field. Recent DB calculations performed by the 
T\"ubingen Group \cite{sefu97,fuchs98} with the Bonn potentials 
would be interesting to study in future applications. 

The out--of--plane dynamics are found to be more difficult to interpret. 
The squeeze--out effect is qualitatively 
reproduced by both models but also substantial differences 
are seen. The azimuthal emission patterns which reflect the transition 
form in--plane to out--of--plane flow in more detail only 
yield a qualitative description of the data. It will be a challenge 
for future investigation to improve on this, most likely by the 
consistent inclusion of in--medium cross sections. However, also here the 
description becomes worse when non-equilibrium effects are neglected 
showing an unrealistic overprediction of the squeeze--out. 

In summary, the present analysis of in--plane and out--of--plane 
observables with microscopic and phenomenological fields and the 
comparison to experimental data can be discussed in two ways: 
Firstly, there are substantial differences in the description between 
using the LDA and LCA approximations, both based on the same DB 
calculations, i.e. on the same EOS. Therefore non--equilibrium effects 
are important in the mean field of heavy ion collisions and the EOS can only 
be extracted reliably by taking these into account. Secondly, the 
calculations with microscopic fields - including non--equilibrium 
effects - describe the data generally well enough, usually better than with 
phenomenological fields. From this we may conclude that microscopic 
nuclear fields can be expected to describe nuclear systems not only 
for nuclear matter and finite nuclei in equilibrium, but also in the 
highly non--equilibrated situation of heavy ion collisions. Thus one can 
expect to move towards a unification of the description of very different 
nuclear systems and to a determination of the EOS of nuclear matter. 
However, further systematic investigations are certainly neccessary to 
corroborate this conclusion.

\begin{ack}
The authors would like to thank the FOPI group for providing us 
with the filter simulation code and the data shown in this work. 
In particular we thank P. Crochet for very helpful comments 
concerning the comparison to the data. Finally we would like to thank 
S. Typel for further helpful discussions.
\end{ack}
\clearpage

\newpage
\begin{table}[hb]
\begin{center}
\begin{tabular}{|l|c|c|c|}
\hline\hline 
                         &  DB       &    NL2     & NL3        \\ \hline\hline
   ${\rm BE}$ $[MeV]$    &   -13.65  &  -16       &    -16     \\ \hline
   $\rho_{sat}$ $[fm^-3]$  &   0.163 &    0.145   &   0.145    \\ \hline
   ${\rm K}^{-1}$ $[MeV]$  &   250   &   200      &   380      \\ \hline
   $m^{\ast}/M$          &   0.58    &   0.83     &   0.70     \\ \hline\hline
\end{tabular}
\end{center}
\caption{\label{sat_tab} Nuclear matter bulk properties for the non--linear 
Walecka model in the NL2 and NL3 
parametrisation \protect\cite{bkm93} and for the microscopic 
DB approach \protect\cite{thm87a}.
 }
\end{table}
\clearpage

\begin{table}[hb]
\begin{center}
\begin{tabular}{|l|ccc|ccc|}
\hline\hline 
          &\multicolumn{3}{c|}{${\overline \Gamma}^{(12)}_s$} 
          &\multicolumn{3}{c|}{${\overline \Gamma}^{(12)}_0$} \\ \hline
$|{\bf v}|/c$  & $\alpha_s$  & $\beta_s$ & $\gamma_s$ & $\alpha_0$ & $\beta_0$ & $\gamma_0$ \\ \hline
0.0   & 14.3824     & 5.9953    & 17.5509   & 10.9222    & 6.1806    & 13.526       \\ 
0.1   & 17.1786     & 5.8245    & 17.7130   & 14.3573    & 6.3442    & 13.3728      \\  
0.2   & 16.9493     & 5.1213    & 17.6029   & 13.1190    & 5.2685    & 13.0947      \\  
0.3   & 16.3484     & 4.4216    & 17.2461   & 12.7487    & 4.4378    & 12.5720      \\  
0.4   & 14.4627     & 3.4204    & 16.5104   & 11.2137    & 3.2912    & 11.7113      \\  
0.5   & 12.9065     & 2.3405    & 15.1360   & 10.0609    & 1.9533    & 9.9478       \\  
0.6   & 12.2228     & 1.8107    & 14.0333   & 9.9916     & 1.3334    & 8.2921       \\  
0.7   & 10.5331     & 2.4902    & 15.4875   & 7.5123     & 2.2559    & 10.4210      \\  
0.8   & 10.5551     & 3.5655    & 16.1264   & 7.3937     & 3.7189    & 11.0364      \\  
0.9   & 11.4589     & 3.9925    & 15.6650   & 8.2575     & 4.5667    & 10.6923      \\  
0.99  & 12.0807     & 4.0514    & 15.0879   & 8.9602     & 4.9836    & 10.2812      \\ \hline\hline
\end{tabular}
\end{center}
\caption{\label{Gamma_12_Tab}
Coefficients $\alpha_{s,0}$, $\beta_{s,0}$ and $ \gamma_{s,0}$ for the 
parameterization, Eq. (\protect\ref{gamma_12_fit}), 
of the effective non-equilibrium DB vertex functions as functions of the c.m. 
nuclear matter streaming velocities}
\end{table}
\clearpage
\begin{table}[hb]
\begin{center}
\begin{tabular}{|l|c|c|c|}
\hline\hline 
$E_{{\rm beam}}$  &  250 A.MeV & 400 A.MeV & 600 A.MeV   \\ \hline\hline
   FOPI     &          44      &    55     &    62       \\ \hline
   NL2      &          52      &    58     &    66       \\ \hline
   DB/LCA   &          50      &    52     &    58       \\ \hline
   DB/LDA   &          51      &    54     &    60       \\ \hline\hline
\end{tabular}
\end{center}
\caption{\label{pm-tab} Lower limits of the highest multiplicity bin PM5 
in Au on Au reactions at various incident energies. The calculations are 
performed with DB forces including (DB/LCA) and without (DB/LDA) 
non-equilibrium effects and with the $\sigma\omega$ 
model (NL2). The corresponding experimental values (FOPI) are 
taken from Ref. \protect\cite{croch1}.
 }
\end{table}
\clearpage
\begin{table}[hb]
\begin{center}
\begin{tabular}{|l|l|c|c|c|}
\hline\hline 
\multicolumn{2}{|c|}{$E_{{\rm beam}}$}   
         &  250 A.MeV      & 400 A.MeV     & 600 A.MeV \\ \hline\hline
                & $<b_{{\rm FOPI}}>$     &   6.3  &  6.3  &  5.5  \\ \cline{2-5}
$PM3$           & $<b_{{\rm NL2}}>$      &   8.8  &  9.0  &  8.5  \\ \cline{2-5}
                & $<b_{{\rm DB/LCA}}>$   &   6.5  &  8.6  &  8.6  \\ \hline\hline
                & $<b_{{\rm FOPI}}>$     &   4.2  &  4.1  &  3.4  \\ \cline{2-5}
$PM4$           & $<b_{{\rm NL2}}>$      &   5.3  &  5.3  &  5.1  \\ \cline{2-5}
                & $<b_{{\rm DB/LCA}}>$   &   3.9  &  5.0  &  4.9  \\ \hline\hline
                & $<b_{{\rm FOPI}}>$     &   1.8  &  1.6  &  1.2  \\ \cline{2-5}
$PM5$           & $<b_{{\rm NL2} }>$     &   2.0  &  1.3  &  2.0  \\ \cline{2-5}
                & $<b_{{\rm DB/LCA} }>$  &   1.3  &  1.3  &  1.3  \\ \hline\hline
\end{tabular}
\end{center}
\caption{\label{bmean-tab}
Correlation between the mean impact parameter $< b >$ [fm] 
and the multiplicity bins in Au on Au reactions at various incident 
energies. The calculations are 
performed with non-equilibrium DB forces (DB/LCA) and 
with the $\sigma\omega$ 
model (NL2). The corresponding experimental values (FOPI) are 
obtained in with sharp cutoff model \protect\cite{crochthesis}.
}
\end{table}
\clearpage
\begin{figure}[t]
\begin{center}
\leavevmode
\epsfxsize=12cm
\epsffile[85 310 520 470]{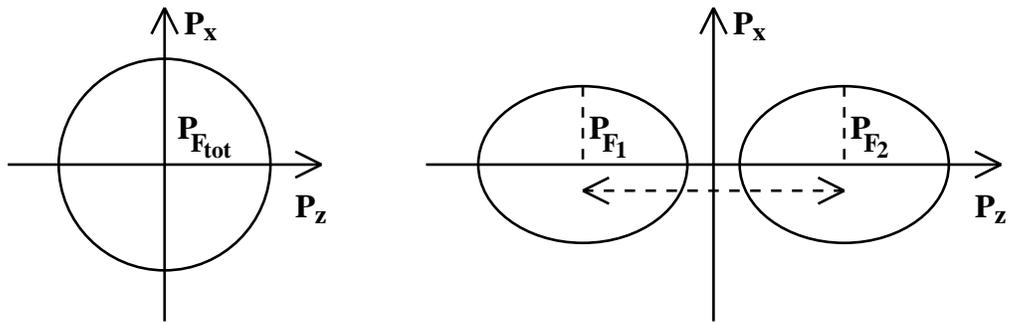}
\end{center}
\caption{\label{LCA_graph}
Schematic representation of the local momentum 
space which corresponds to the
local density approximation (left) 
and the local configuration approximation (right).
} 
\end{figure}
\begin{figure}[b]
\begin{center}
\leavevmode
\epsfxsize = 13cm
\epsffile[60 41 389 388]{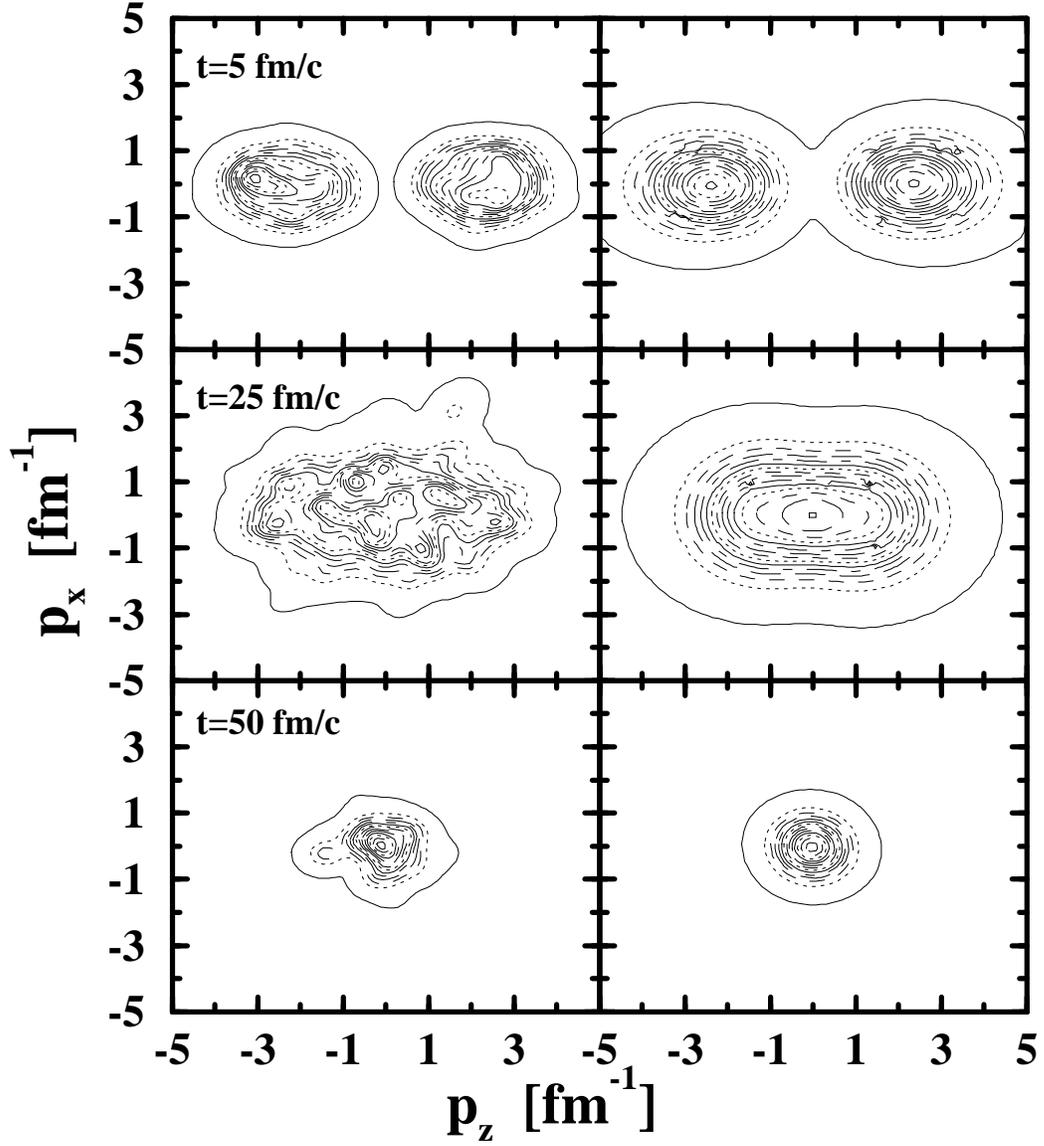}
\end{center}
\caption{\label{phasespace_graph}
Local momentum distributions obtained from the transport calculation 
(left columns) 
and the corresponding fitted two hot Fermi--sphere distributions 
(right columns) 
at different stages ($t=5,25$ and $50$ fm/c) in the central region 
of a central ($b=0$ fm) Au+Au reaction at $600$ A.MeV.
}
\end{figure}
\begin{figure}[b]
\begin{center}
\leavevmode
\epsfxsize = 13cm
\epsffile[30 85 430 410]{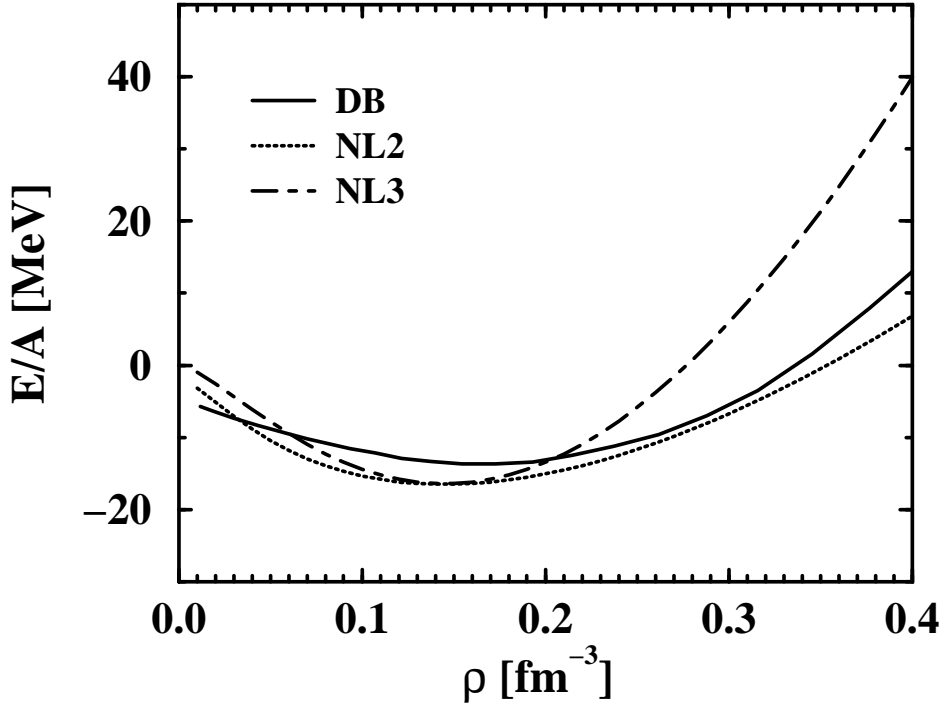}
\end{center}
\caption{\label{EOS_graph}
Equation--of--state in the DB approach \protect\cite{thm87a} 
and in the non--linear $\sigma\omega$--model with parameter sets 
NL2 (soft) and NL3 (hard).
}
\end{figure}
\begin{figure}[b]
\begin{center}
\begin{tabular}{cc}
                \leavevmode
                \epsfxsize = 7.5cm
                \epsffile[155 240 450 450]{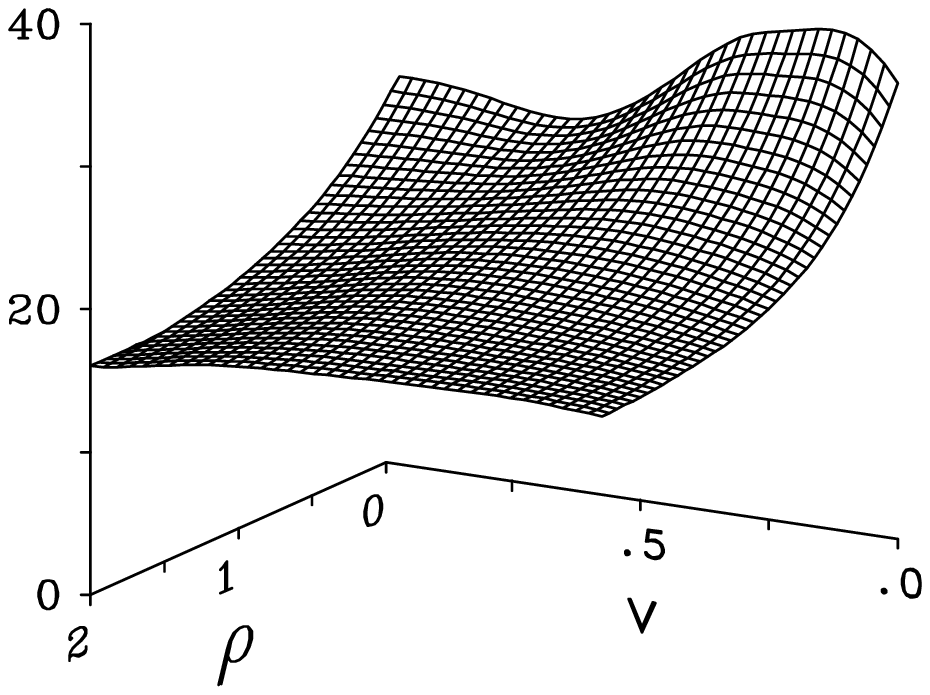}
                &
                \leavevmode
                \epsfxsize = 7.5cm
                \epsffile[155 270 450 480]{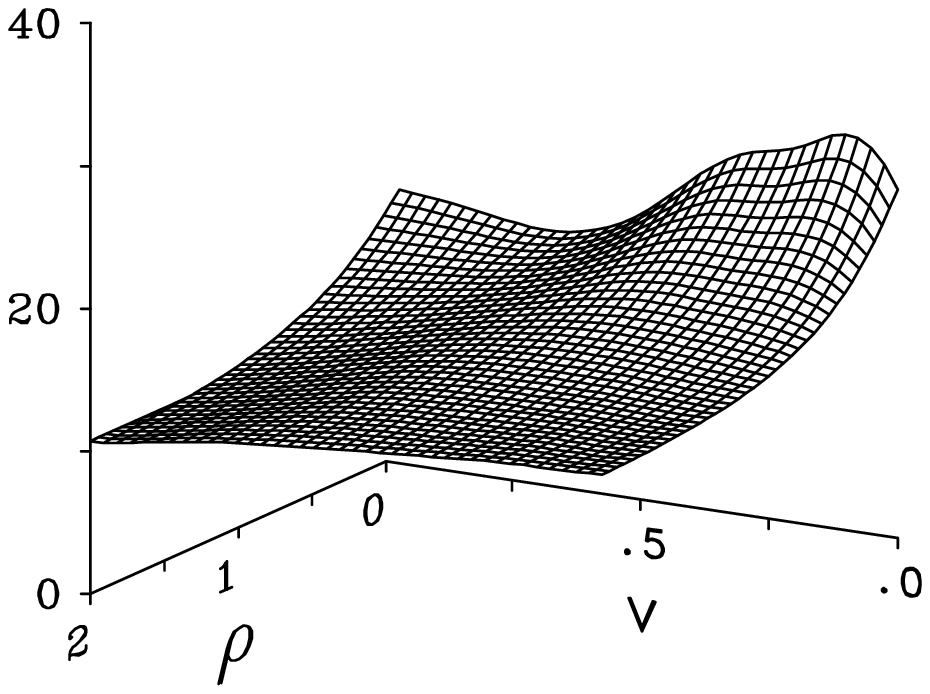}
\end{tabular}
\end{center}
\caption{\label{gamma_12_graph}
Dynamical vertex functions $ {\overline \Gamma}^{(12)}_{s}$ (left) and 
$ {\overline \Gamma}^{(12)}_{0}$ (right) in symmetric 
colliding nuclear matter as functions of the subsystem 
density $\rho_{0}^{(i)}$ and the streaming 
velocity $v$ of the subsystems 
in dimensionless units ($\times\frac{M^2}{4\pi}$).
}
\end{figure}
\begin{figure}[b]
\begin{center}
\leavevmode
\epsfxsize = 13cm
\epsffile[30 85 430 410]{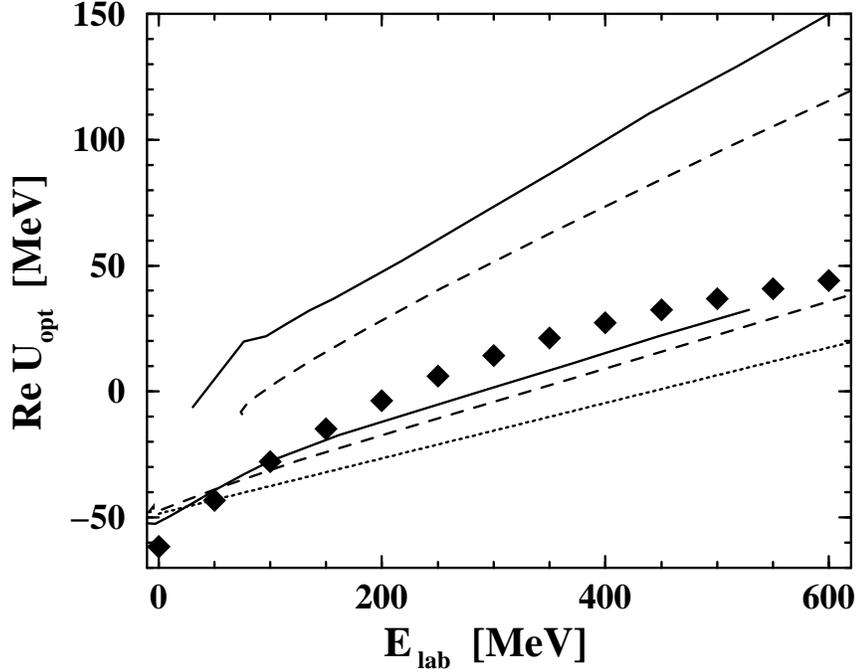}
\end{center}
\caption{\label{opt_graph}
Energy dependence of different optical potentials. The solid lines 
represent the DB nucleon--nucleus optical potential 
\protect\cite{thm87a} at saturation density $\rho_{{\rm sat}}$ 
(lower curve) and 2$\rho_{{\rm sat}}$ (upper curve). 
The dashed lines represent the 
corresponding nucleon optical potential in a nucleus--nucleus 
collision determined in the colliding nuclear matter 
approximation at subsystem densities 
$\rho_{0}^{(1)}+\rho_{0}^{(2)} =\rho_{{\rm sat}}$ (lower curve) and 
$\rho_{0}^{(1)}+\rho_{0}^{(2)} =2 \rho_{{\rm sat}}$ (upper curve). 
The dotted line refers to the nucleon--nucleus optical 
potential at $\rho_{sat}$ in the non--linear $\sigma\omega$--model NL2, 
and the diamonds are the corresponding empirical values 
\protect\cite{hama90}.
}
\end{figure}
\begin{figure}[b]
\begin{center}
\leavevmode
\epsfxsize = 13cm
\epsffile[0 90 530 580]{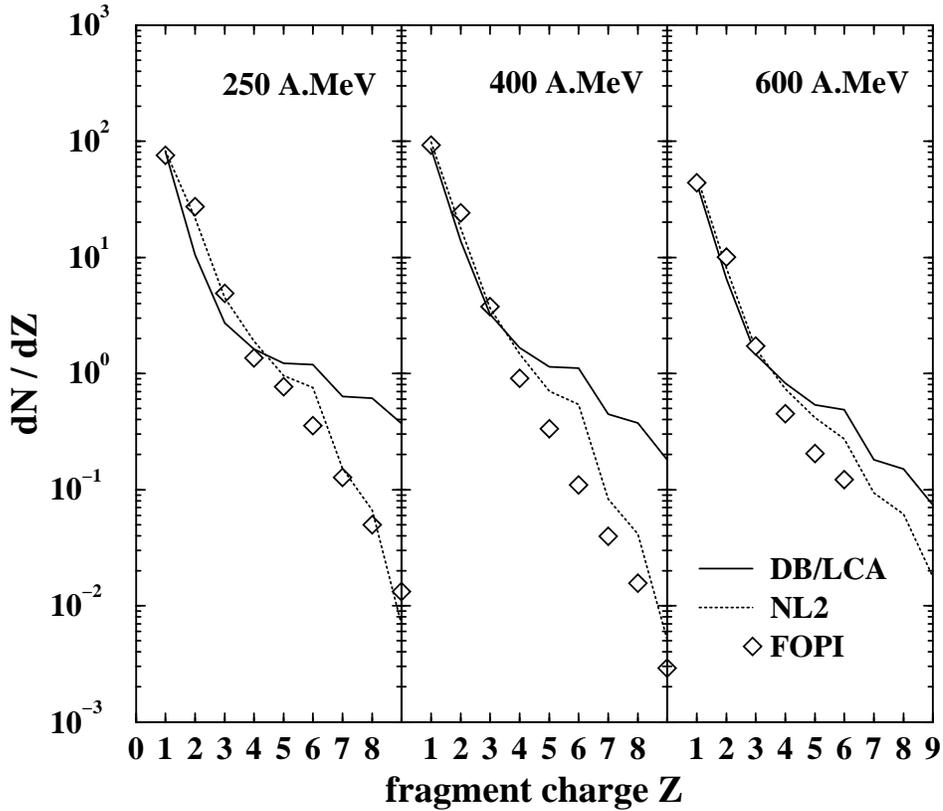}
\end{center}
\caption{\label{fig4.2} 
Comparison of the theoretical charge distributions to 
the FOPI data \protect\cite{reis97} for central  
Au+Au reactions at 250 (left), 400 (middle) and 600 A.MeV 
(right). The solid and dotted lines represent the 
charge distributions determined in the 
DB/LCA  and NL2 models, respectively. 
At $250$ and $400$ A.MeV the centrality class was determined 
by the $ERAT$ selection with the corresponding $ERAT$--bins taken from 
Ref. \protect\cite{reis97}. At $600$ A.MeV the $PM4$ selection was used 
(see section 6.1).
}
\end{figure}
\begin{figure}[b]
\begin{center}
\leavevmode
\epsfxsize = 13cm
\epsffile[0 90 530 580]{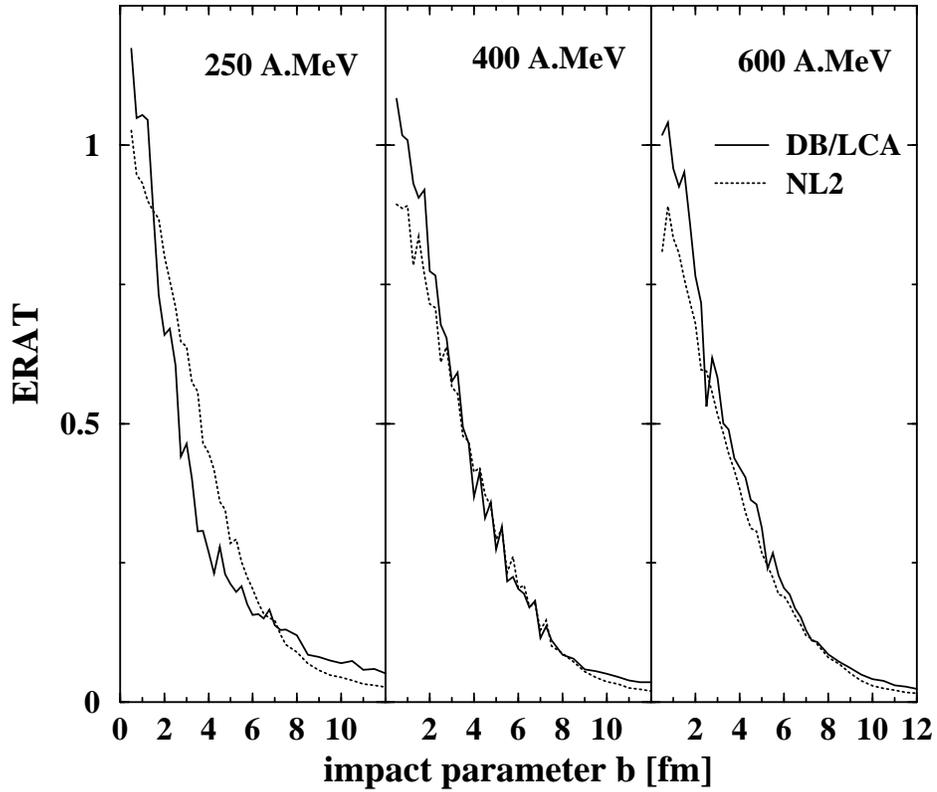}
\end{center}
\caption{\label{fig5.1a} 
Correlation between the observable $ERAT$ and the impact parameter for 
Au+Au reactions at 250 (left), 400 (middle) and 600 A.MeV 
(right). The curves have the same meaning as in 
Fig. \protect\ref{fig4.2}. 
}
\end{figure}
\begin{figure}[b]
\begin{center}
\leavevmode
\epsfxsize = 13cm
\epsffile[0 90 530 580]{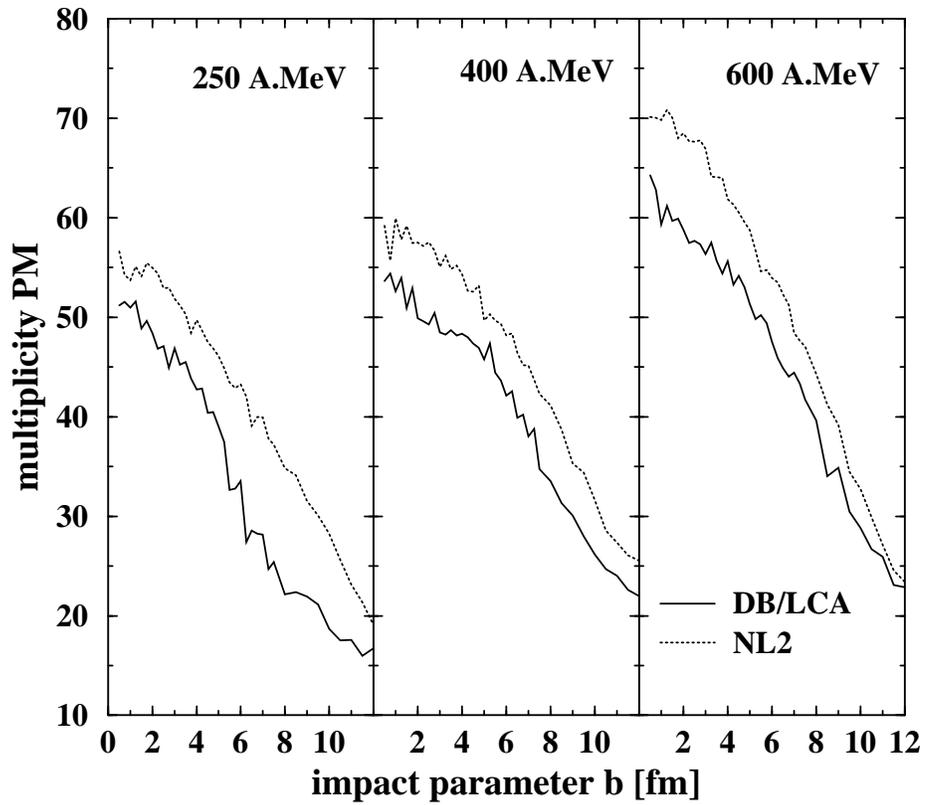}
\end{center}
\caption{\label{fig5.1b} 
Correlation between the mean multiplicity $PM$ and the 
impact parameter for 
Au+Au reactions at 250 (left), 400 (middle) and 600 A.MeV 
(right). The curves have the same meaning as in 
Fig. \protect\ref{fig4.2}. 
}
\end{figure}
\begin{figure}[b]
\begin{center}
\leavevmode
\epsfxsize = 13cm
\epsffile[35 160 450 570]{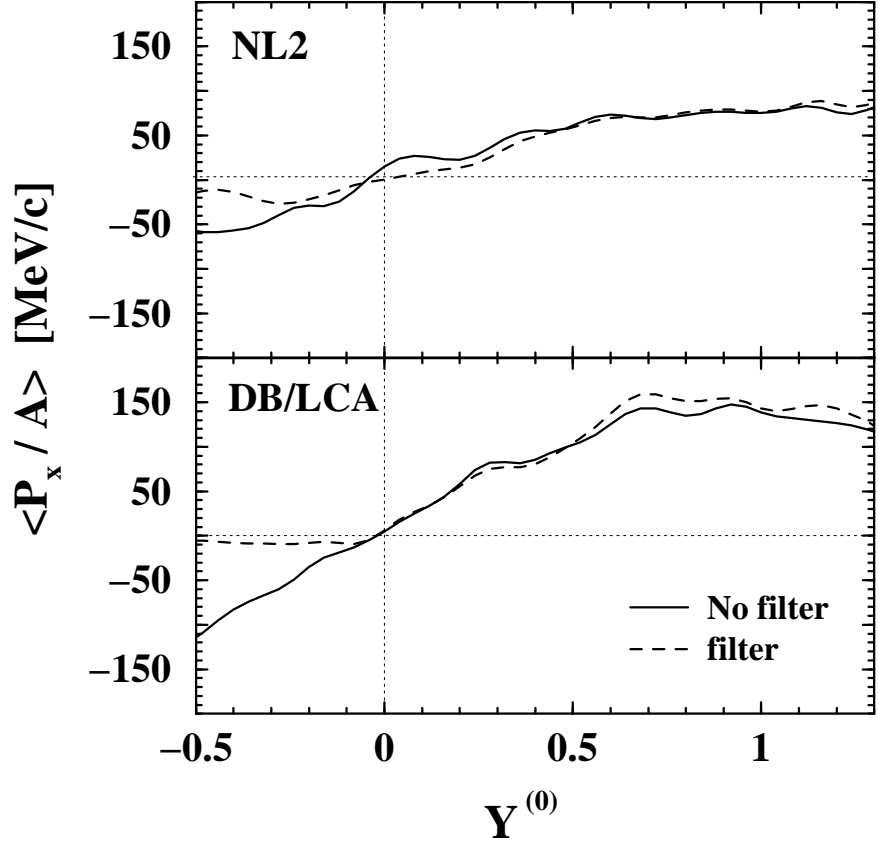}
\end{center}
\caption{\label{fig5.2a} 
Mean in--plane flow $<p_{x}/A>$ in a semi-central 
($b=3$ fm) Au+Au reaction at $600$ A.MeV. 
The model dependence and the influence of the FOPI filter 
on the sideward flow is shown. The solid lines represent 
the unfiltered, the dashed lines the filtered results 
obtained in the NL2 (top) and DB/LCA (bottom) model, respectively.
}
\end{figure}
\begin{figure}[b]
\begin{center}
\leavevmode
\epsfxsize = 13cm
\epsffile[36 26 579 339]{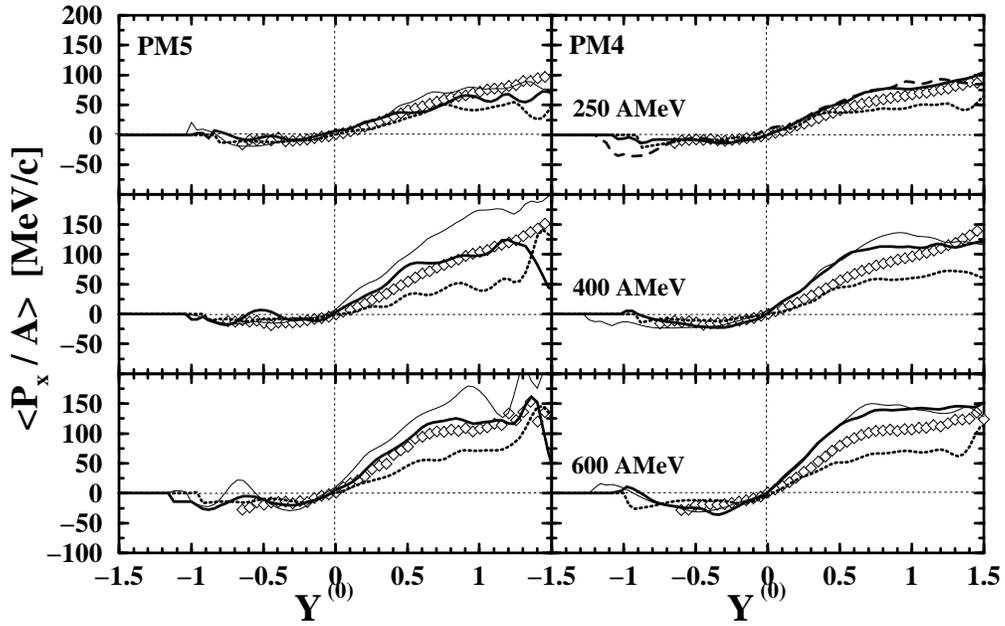}
\end{center}
\caption{\label{fig5.2b} 
Comparison of the sideward flow of 
protons ($Z=1$) with the FOPI data (diamonds) taken from 
\protect\cite{crochthesis,croch3}. The calculations are 
performed in the DB model with (DB/LCA, solid) and 
without non-equilibrium effects (DB/LDA, dashed) and 
in the non-linear $\sigma\omega$ model NL2 (dotted). 
The left panels correspond to central ($PM5$), 
the right panels to peripheral ($PM4$) Au on Au reactions at 
different beam energies of $250$ (top), $400$ (middle) and 
$600$ (bottom) A.MeV.
}
\end{figure}
\begin{figure}[b]
\begin{center}
\leavevmode
\epsfxsize = 13cm
\epsffile[36 26 579 339]{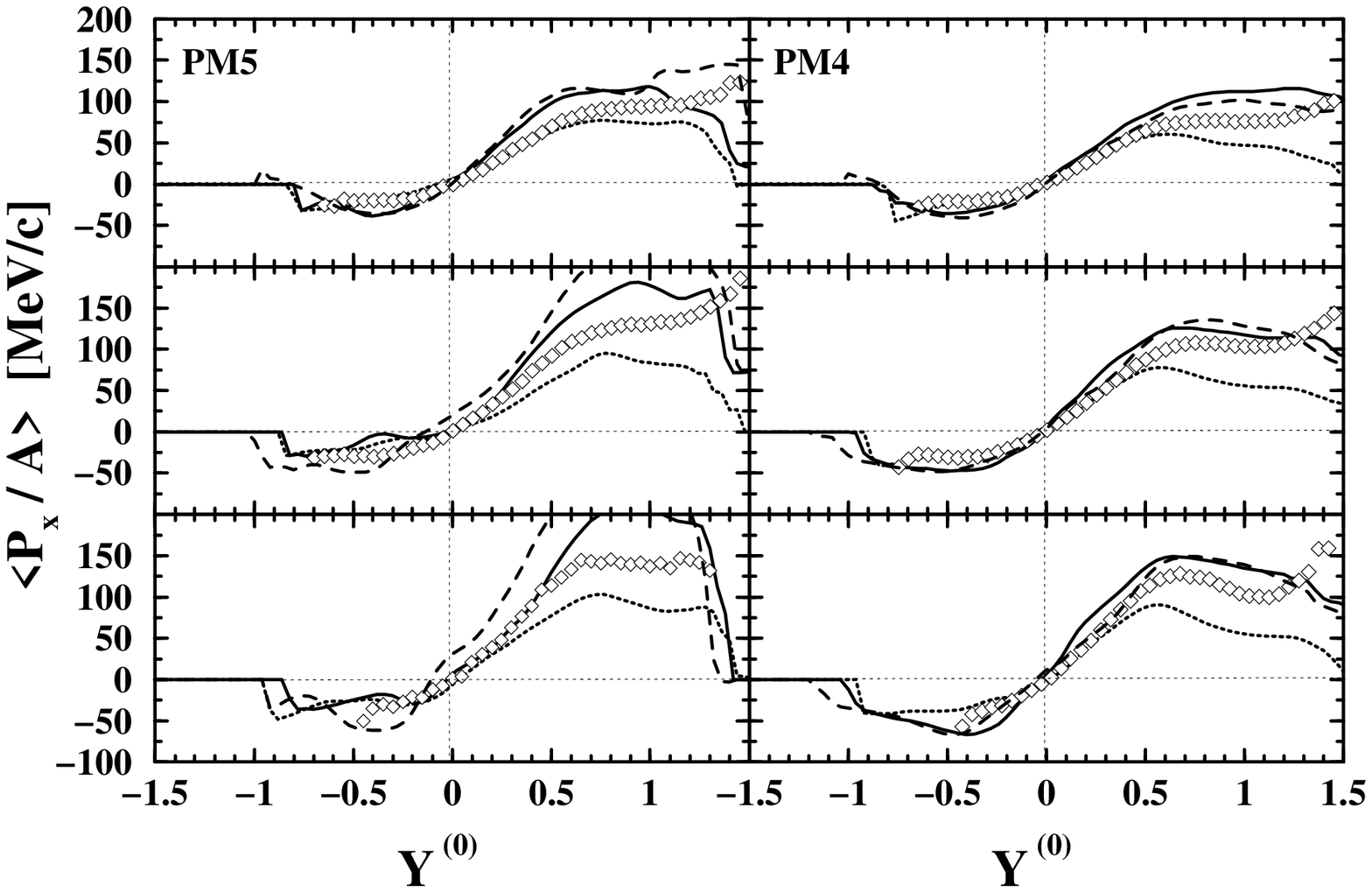}
\end{center}
\caption{\label{fig5.2bz2} 
Comparison of the (filtered) sideward flow per 
nucleon of fragments with charge $Z=2$ with the FOPI data (diamonds) 
taken from \protect\cite{crochthesis,croch3}. 
The curves have the same meaning as 
in Fig. \protect\ref{fig5.2b}.
}
\end{figure}
\begin{figure}[b]
\begin{center}
\leavevmode
\epsfxsize = 13cm
\epsffile[36 26 579 339]{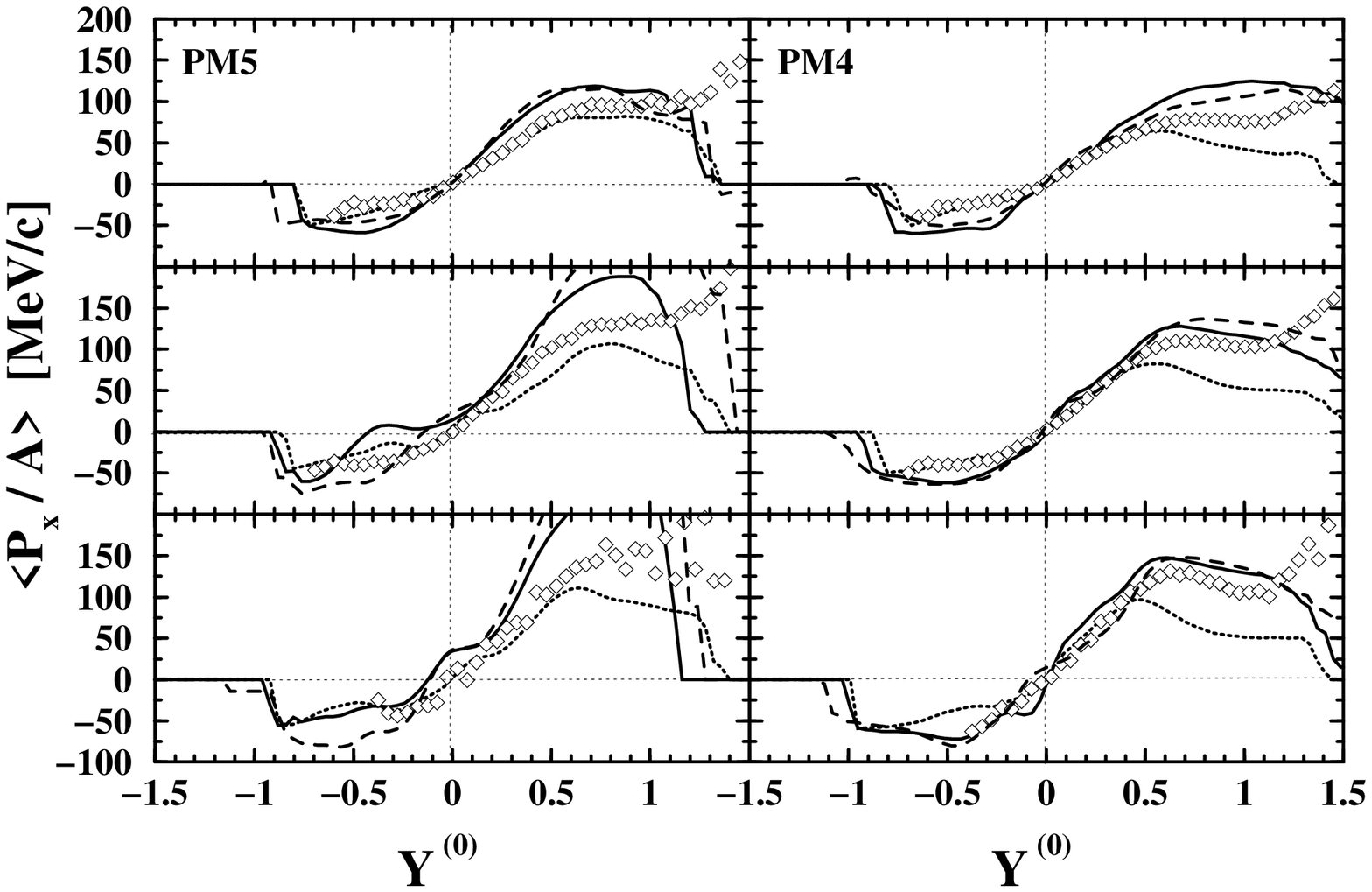}
\end{center}
\caption{\label{fig5.2bz3} 
Comparison of the sideward flow per 
nucleon of fragments with charge $Z=3$ with the FOPI data (diamonds) 
taken from \protect\cite{crochthesis,croch3}. 
The curves have the same meaning as 
in Fig. \protect\ref{fig5.2b}.
}
\end{figure}
\begin{figure}[b]
\begin{center}
\leavevmode
\epsfxsize = 13cm
\epsffile[36 26 579 339]{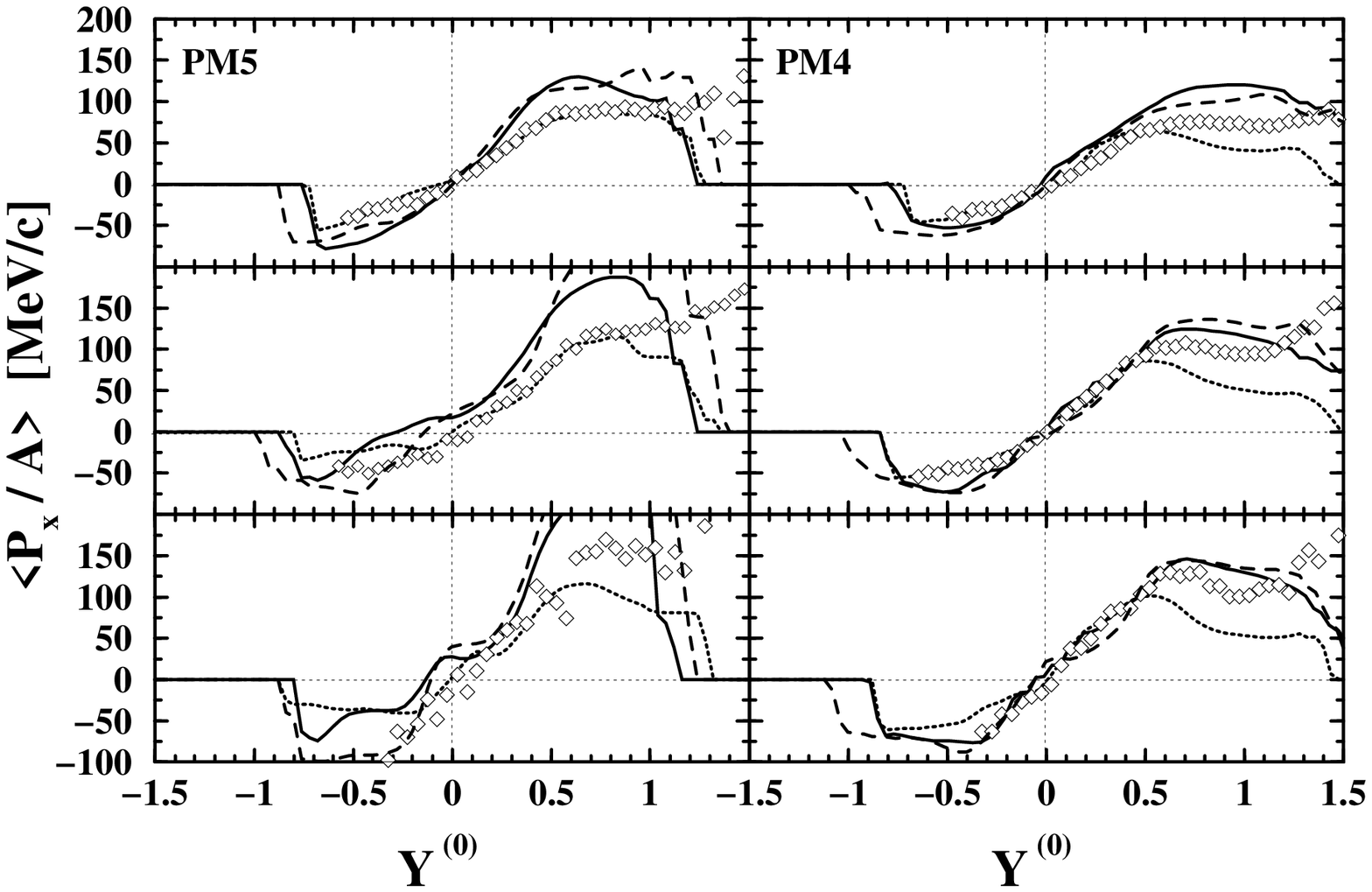}
\end{center}
\caption{\label{fig5.2bz4} 
Comparison of the sideward flow per 
nucleon of fragments with charge $Z=4$ with the FOPI data (diamonds) 
taken from \protect\cite{crochthesis,croch3}. The curves have 
the same meaning as in Fig. \protect\ref{fig5.2b}.
}
\end{figure}
\clearpage
\begin{figure}[b]
\begin{center}
\leavevmode
\epsfxsize = 13cm
\epsffile[77 66 433 245]{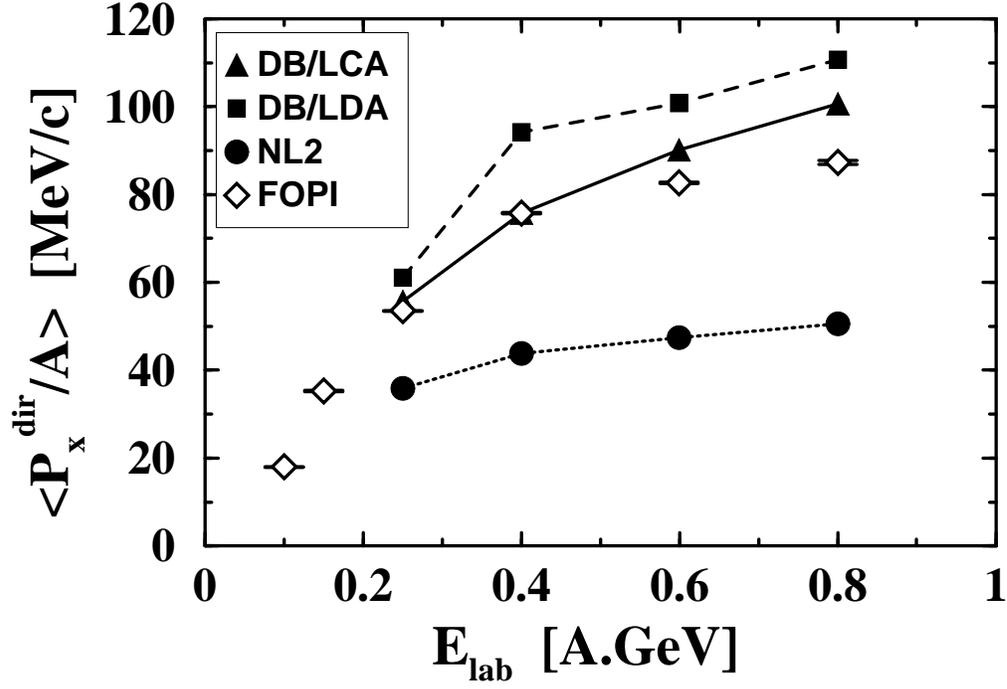}
\end{center}
\caption{\label{fig5.2c} 
Mean directed in--plane sideward flow per nucleon in semi--central ($PM4$) 
Au on Au reactions as a function of the beam energy. Calculations 
performed with DB model including (DB/LCA, triangles) and 
without non-equilibrium effects (DB/LDA, squares) and 
with the non-linear $\sigma\omega$ model NL2 (circles)
are compared to the FOPI data 
(diamonds) taken from Ref. \protect\cite{crochthesis,croch3}.
}
\end{figure}
\begin{figure}[b]
\begin{center}
\leavevmode
\epsfxsize = 15cm
\epsffile[26 47 596 462]{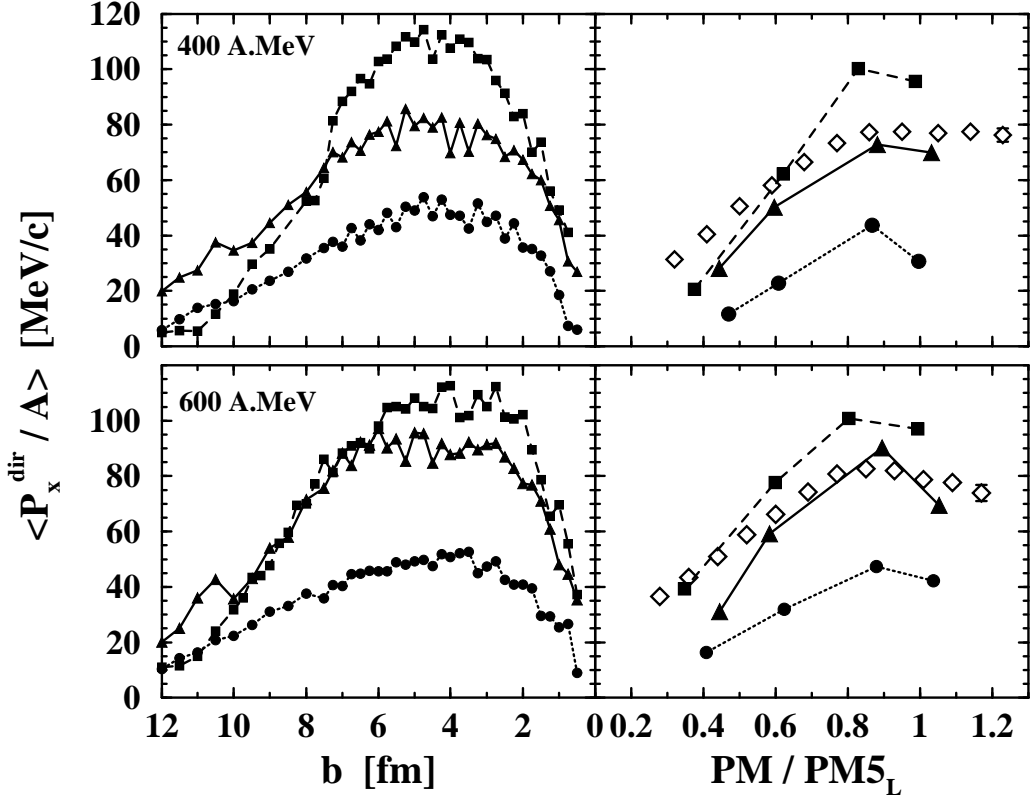}
\end{center}
\caption{\label{fig5.2d} 
Centrality dependence of the mean directed in-plane 
flow per nucleon for the system Au on Au at $400$ (top panels) 
and $600$ (bottom panels) A.MeV. The left panels show the impact 
parameter dependence of the calculation; on the right panels the 
calculations using 
the multiplicity selection are compared to the FOPI data 
\protect\cite{crochthesis,croch3}. Triangles refer to DB/LCA, 
squares to DB/LDA and circles to NL2.
}
\end{figure}
\begin{figure}[b]
\begin{center}
\leavevmode
\epsfxsize = 13cm
\epsffile[19 34 534 413]{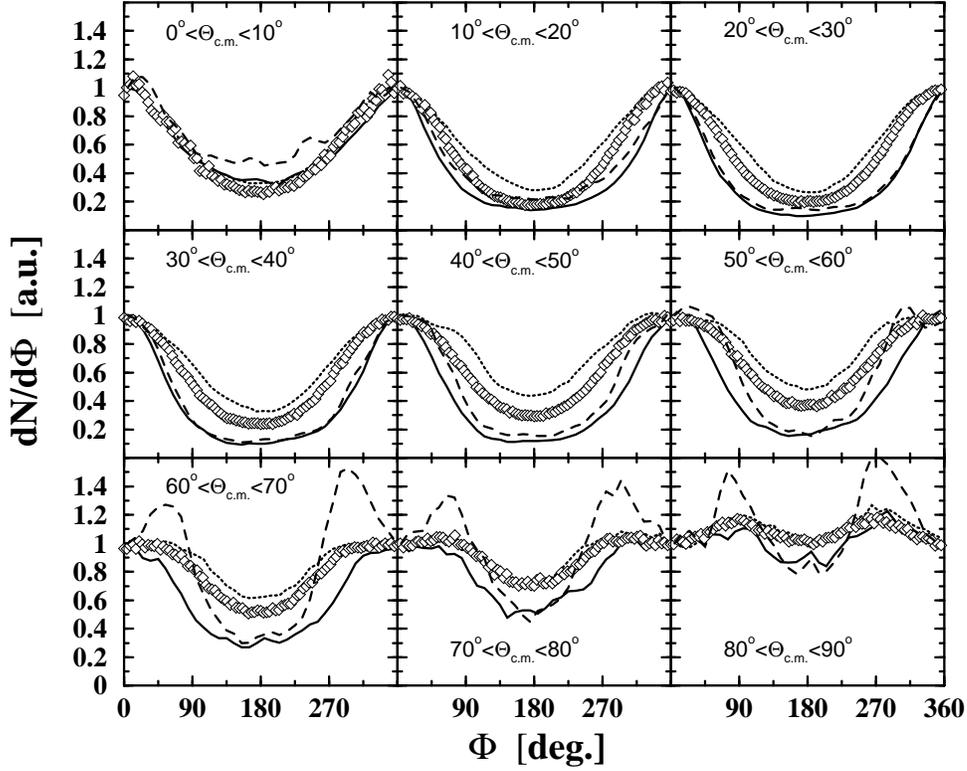}
\end{center}
\caption{\label{fig5.3a} 
Azimuthal distributions at different polar angles 
$\Theta_{{\rm c.m.}}$ for a semi-central Au on Au 
reaction at $600$ A.MeV. The calculations are 
performed within the DB model including (DB/LCA, solid) 
and without (DB/LDA, dashed) non-equilibrium effects 
and the non-linear $\sigma\omega$ model NL2 (dotted). 
The diamonds represent the data 
taken from Ref. \protect\cite{crochet97}.
}
\end{figure}
\clearpage
\begin{figure}[b]
\begin{center}
\leavevmode
\epsfxsize = 13cm
\epsffile[77 66 433 245]{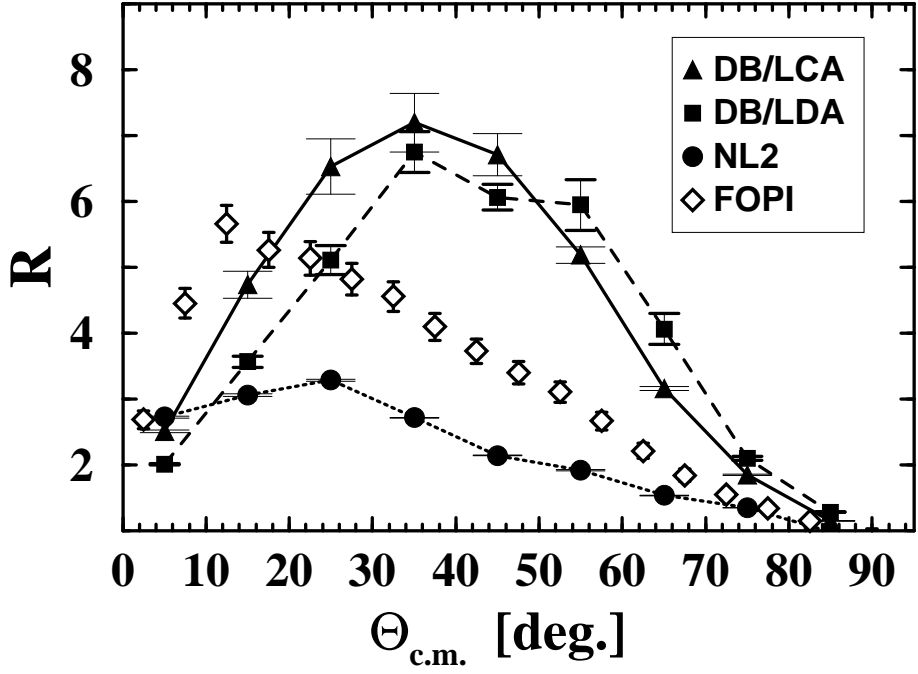}
\end{center}
\caption{\label{fig5.3c} 
Dependence of the in-plane emission, i.e. the 
azimuthal anisotropy ratio $R$, Eq. (\protect\ref{ratio}), on the 
c.m. polar angle in a semi-central Au on Au reaction at 600 A.MeV. 
The calculations are 
performed within the DB model including (DB/LCA, solid) 
and without (DB/LDA, dashed) non-equilibrium effects 
and the non-linear $\sigma\omega$ model NL2 (dotted). 
The diamonds refer to the FOPI data of Ref. \protect\cite{crochet97}. 
}
\end{figure}
\clearpage
\begin{figure}[b]
\begin{center}
\leavevmode
\epsfxsize = 13cm
\epsffile[71 26 490 416]{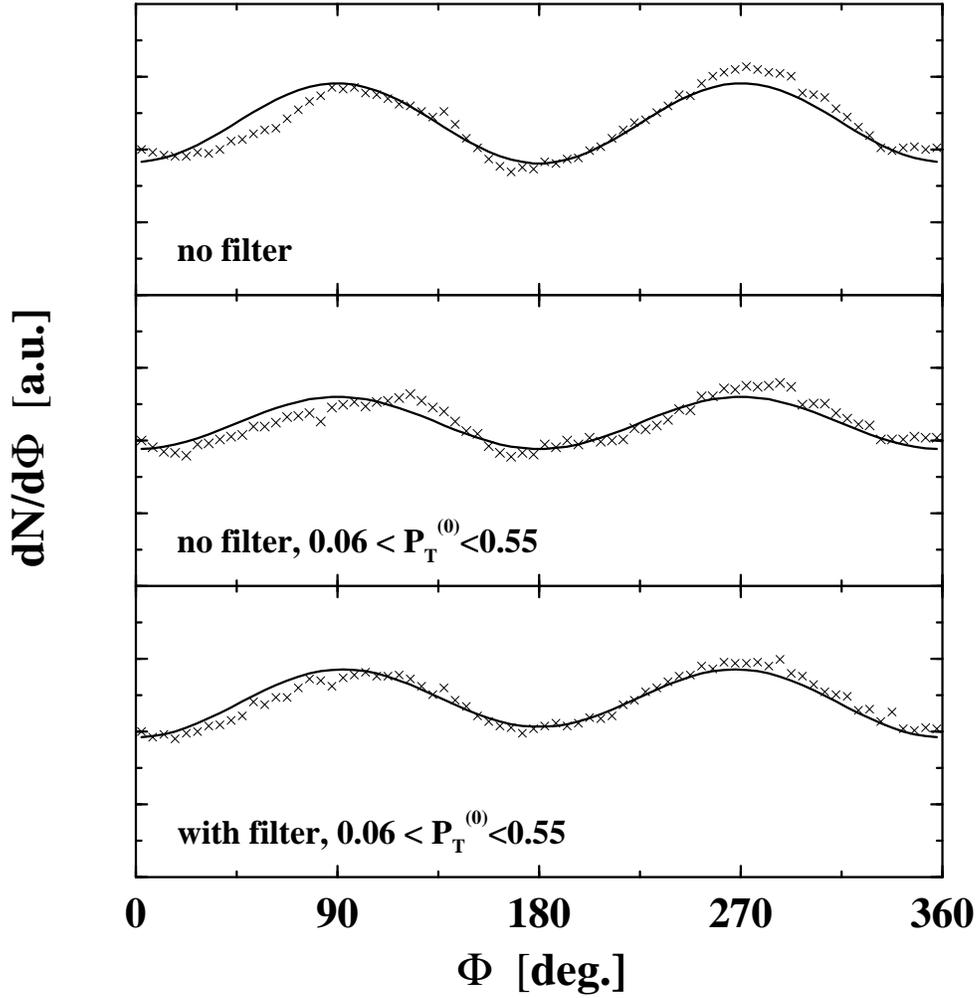}
\end{center}
\caption{\label{fig5.3e} 
Influence of the FOPI filter on the azimuthal distributions for a 
semi-central ($PM4$) Au on Au reaction at $600$ A.MeV. 
The calculations are performed with the NL2 model. 
The solid lines are fits to the curves according 
to Eq. (\protect\ref{fit}).
}
\end{figure}
\begin{figure}[b]
\begin{center}
\leavevmode
\epsfxsize = 13cm
\epsffile[90 10 490 360]{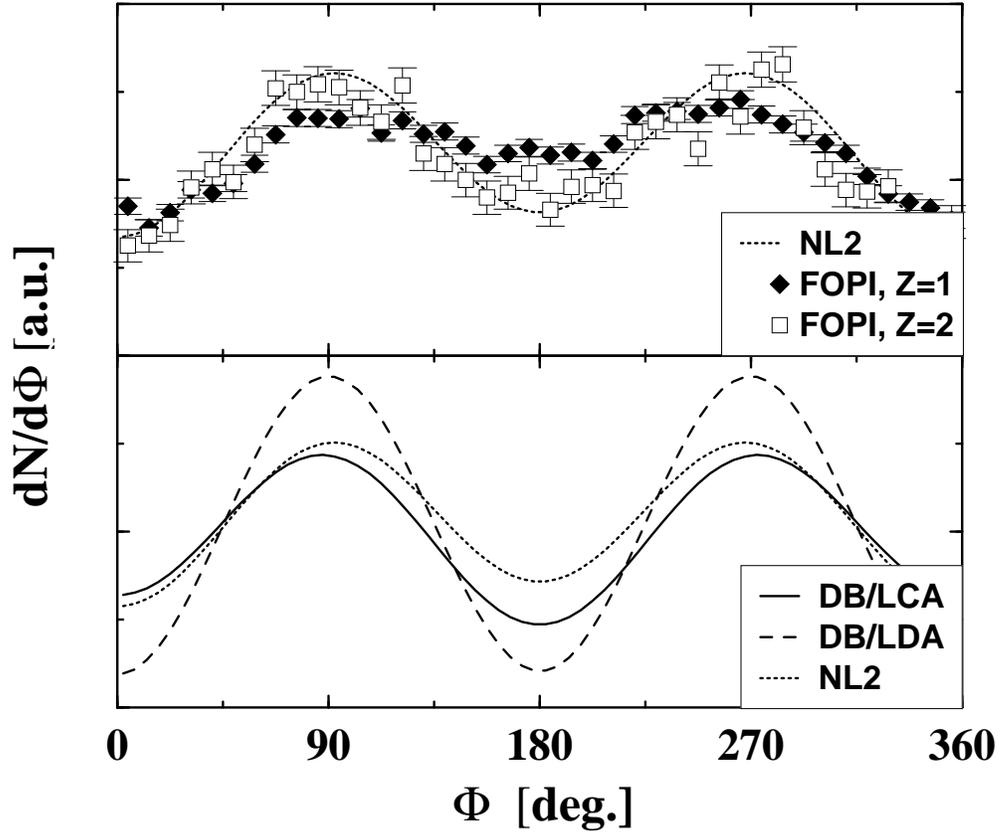}
\end{center}
\caption{\label{fig5.3f} 
Lower panel: Azimuthal distributions of protons ($Z=1$) 
in a semi-central ($PM4$) 
Au on Au reaction at $600$ A.MeV.
The calculations are performed with the DB/LCA, DB/LDA and the 
NL2 forces. 
Upper panel: The NL2 results are compared to the 
corresponding FOPI data \protect\cite{croch2} for $Z=1$ and $Z=2$. 
}
\end{figure}
\begin{figure}[b]
\begin{center}
\leavevmode
\epsfxsize = 13cm
\epsffile[77 66 433 245]{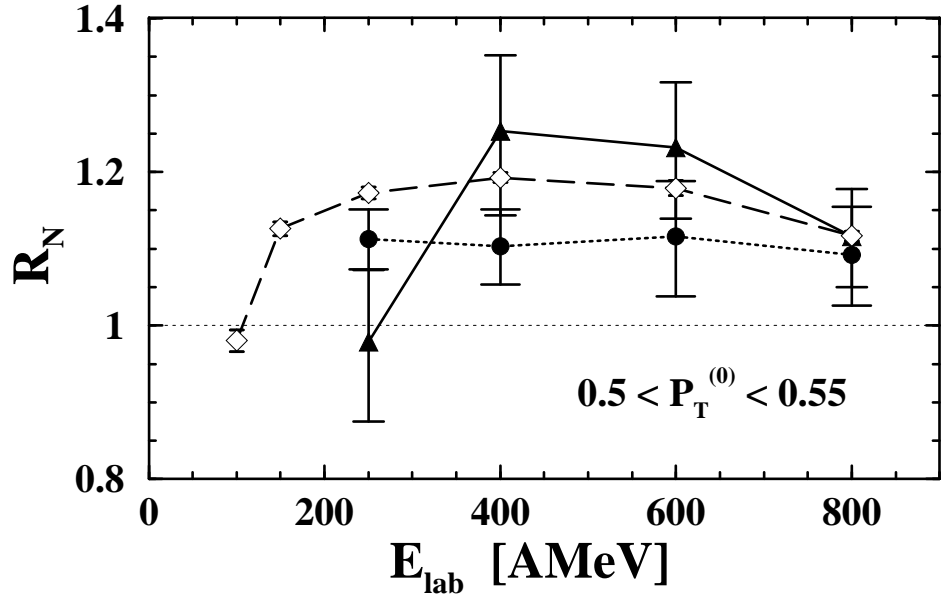}
\end{center}
\caption{\label{fig5.3g} 
Squeeze-out ratio $R_{N}$ as a function of the incident energy 
for semi-central ($PM4$) Au on Au reactions with a 
$0.5 \leq P_{T}^{(0)} \leq 0.55$ cut. The theoretical 
calculations for the NL2 (circles) and the non-equilibrium DB 
forces (triangles) are shown and compared to the FOPI 
data (open diamonds) taken from Ref. \protect\cite{crochthesis}.  
}
\end{figure}
\end{document}